\documentclass[12pt]{article}

\usepackage{graphicx}

\def\be{\begin{equation}}
\def\ee{\end{equation}}
\def\ba{\begin{eqnarray}}
\def\ea{\end{eqnarray}}
\def\ge{\mathrel{\raise.3ex\hbox{$>$\kern-.75em\lower1ex\hbox{$\sim$}}}}
\def\la{\mathrel{\raise.3ex\hbox{$<$\kern-.75em\lower1ex\hbox{$\sim$}}}}

\def\simgt{\mathrel{\raise.3ex\hbox{$>$\kern-.75em\lower1ex\hbox{$\sim$}}}}
\def\simlt{\mathrel{\raise.3ex\hbox{$<$\kern-.75em\lower1ex\hbox{$\sim$}}}}

\newcommand{\bi}[1]{\bibitem{#1}}
\newcommand{\fr}[2]{\frac{#1}{#2}}

\newcommand{\tb}{\tan\beta}

\newcommand{\nc}{\newcommand}

\nc{\gone}{\bar g_{\pi NN}^{(1)}}
\nc{\gzero}{\bar g_{\pi NN}^{(0)}}
\nc{\al}{\alpha}
\nc{\ga}{\gamma}
\nc{\de}{\delta}
\nc{\ep}{\epsilon}
\nc{\ze}{\zeta}
\nc{\et}{\eta}
\renewcommand{\th}{\theta}
\nc{\Th}{\Theta}
\nc{\ka}{\kappa}
\nc{\rh}{\rho}
\nc{\si}{\sigma}
\nc{\ta}{\tau}
\nc{\up}{\upsilon}
\nc{\ph}{\phi}
\nc{\ch}{\chi}
\nc{\ps}{\psi}
\nc{\om}{\omega}
\nc{\Ga}{\Gamma}
\nc{\De}{\Delta}
\nc{\La}{\Lambda}
\nc{\Si}{\Sigma}
\nc{\Up}{\Upsilon}
\nc{\Ph}{\Phi}
\nc{\Ps}{\Psi}
\nc{\Om}{\Omega}
\nc{\ptl}{\partial}
\nc{\del}{\nabla}
\nc{\ov}{\overline}
\nc{\newcaption}[1]{\centerline{\parbox{15cm}{\caption{#1}}}}

\begin{document}

\begin{titlepage}

\rightline{UMN--TH--2401/05}
\rightline{FTPI--MINN--05/13}
\rightline{CERN-PH-TH-2005-095}
\rightline{hep-ph/0506106}

\setcounter{page}{1}

\vspace*{0.2in}

\begin{center}

\hspace*{-0.6cm}\parbox{17.5cm}{\Large \bf \boldmath{$CP$}-odd Phase Correlations and Electric Dipole Moments}

\vspace*{0.5cm}
\normalsize

{\bf Keith A. Olive$^{\,(a)}$, Maxim Pospelov$^{\,(b,c,d)}$, 
Adam Ritz$^{\,(e)}$ and Yudi Santoso$^{\,(b,c)}$}

\smallskip
\medskip

$^{\,(a)}${\it W. I. Fine Theoretical Physics Institute, School of Physics and
Astronomy,\\  University of Minnesota, Minneapolis, MN 55455, USA}

$^{\,(b)}${\it Department of Physics, University of Guelph, Guelph,\\
Ontario N1G 2W1, Canada}

$^{\,(c)}${\it Perimeter Institute for Theoretical Physics, Waterloo,
Ontario N2J 2W9, Canada}

$^{\,(d)}${\it Department of Physics and Astronomy, University of Victoria, \\
     Victoria, BC, V8P 1A1 Canada}

$^{\, (e)}${\it Theoretical Division, Department of Physics, CERN,
Geneva 23, CH-1211 Switzerland}
\smallskip
\end{center}
\vskip0.2in

\centerline{\large\bf Abstract}

We revisit the constraints imposed by electric dipole moments (EDMs) of nucleons and heavy atoms
on new $CP$-violating sources within supersymmetric theories. We point
out that certain two-loop renormalization group corrections induce significant mixing between
the basis-invariant $CP$-odd phases. In the framework of the constrained minimal supersymmetric 
standard model (CMSSM), the $CP$-odd invariant related to the soft trilinear $A$-phase at the 
GUT scale, $\theta_A$, induces non-trivial and distinct $CP$-odd phases for the three 
gaugino masses at the weak scale. The latter give one-loop contributions to EDMs enhanced 
by $\tan\beta$, and can provide the dominant contribution to the electron 
EDM induced by $\theta_A$. 
We perform a detailed analysis of the EDM constraints within the CMSSM, 
exhibiting the reach, in terms of sparticle spectra, which may be 
obtained assuming generic phases, as well as the limits on the $CP$-odd phases 
for some specific parameter points where detailed phenomenological studies are available. 
We also illustrate how this reach will expand with results from the next generation of experiments
which are currently in development. 

\vfil
\leftline{June 2005}

\end{titlepage}

\section{Introduction}

Electric dipole moments (EDMs) of the neutron \cite{n} and heavy 
atoms and molecules \cite{Tl,Hg,TlF,Xe,Cs,YbF,PbO} are primary  
observables in probing for sources of flavor-neutral $CP$ violation. 
The high degree of precision with which various experiments have put 
limits on possible EDMs translates into stringent constraints on a variety of 
extensions of the Standard Model at and above the electroweak scale (see, e.g. \cite{KL}). 
Currently, the strongest constraints on $CP$-violating parameters arise 
from the atomic EDMs of thallium \cite{Tl} and mercury \cite{Hg}, and that of 
the neutron \cite{n}:
\ba
|d_{\rm Tl}| &<& 9 \times 10^{-25} e\, {\rm cm} \nonumber\\
|d_{\rm Hg}| &<& 2   \times 10^{-28}  e\, {\rm cm}     \\
|d_n|  &<&   6\times 10^{-26} e\, {\rm cm}.\nonumber
\label{explimit}
\ea
When interpreted as a quantity induced purely by the electron EDM $d_e$, the measurement of
$d_{\rm Tl}$ can be translated into a tight bound, 
$|d_e|<1.6\times 10^{-27} e\, {\rm cm} $. 

While the Standard Model CKM paradigm has received remarkable overall confirmation
with the observation of $CP$-violation in the mixing and decays of neutral $B$-mesons,
the motivation for anticipating new sources of $CP$-violation is undiminished. This
is in part through the need to explain baryogenesis, and also through the generic
appearance of new $CP$-violating phases in models of new physics introduced to stabilize the
Higgs sector. In particular, supersymmetric models with minimal field content, i.e. the MSSM, 
allow for the presence of several $CP$-violating phases even in the most restrictive 
ansatz of flavor universality in the squark and slepton sectors. 
The null experimental EDM results pose a serious problem for the 
MSSM with superpartner masses around the (natural) electroweak 
scale. Indeed, a typical $CP$-violating SUSY phase of order one combined with 
${\cal O}(100{\rm~ GeV})$ masses for the superpartners would violate 
experimental constraints by up to three orders of magnitude \cite{WWI}. 
Moreover, ongoing and planned EDM experiments are aiming to improve the level of sensitivity to 
underlying $CP$-odd sources by several additional orders of magnitude.  

Over the years the SUSY $CP$ problem has led to a number of suggestions for how this 
tuning might be alleviated:

\begin{itemize}
\item {\bf small phases}: The simplest solution 
in principle is of course just to demand that the diagonal phases are small (or zero). However, 
a framework with small phases would run counter to the natural 
Standard Model interpretation of $CP$-violation in the mixing and decays of $B$-mesons, which 
requires no analogous tuning. Of course, such a scenario is possible, e.g. in the framework of 
spontaneous $CP$-violation \cite{spontCP1,spontCP2}, but requires a significant amount of 
model building, particularly in terms of constraints on the flavor structure, in order to suppress 
the flavor-diagonal phases while allowing the effective CKM phase to be order one. 
\item {\bf flavor off-diagonal \boldmath{$CP$}-violation}:
As a principle to motivate small diagonal phases, one may interpret the EDM constraints,
in concert with the success of the CKM paradigm, as implying that certain symmetries should ensure
that all $CP$-violation arises via an interplay with flavor structures. This is an appealing
viewpoint, and can be taken as motivation to try and divorce $CP$-violation from the SUSY-breaking
sector. However, even in the case that the soft terms are real, any significant non-universality in 
these terms, i.e. misalignment between the $A$ terms and the Yukawas, will generically lead to
large EDMs via the large $CP$-violating phases in the Yukawa matrices (see e.g. \cite{Abel:2001cv}). 
Thus, even within this seemingly rather restrictive scenario, further constraints must be imposed 
either on the flavor sector \cite{edmflav} or the SUSY-breaking sector to lessen the 
impact from the EDM constraints.
\item {\bf cancellations}: The MSSM contains in principle many new $CP$-violating phases, and while
a lot of these are highly constrained by the flavor sector, there are still
several new flavor-diagonal phases. Indeed, there are more phases than experimental EDM observables
with which to bound them, and thus various partial cancellations and degeneracies are to be 
expected \cite{WWII,fo2,IN,WWIIa}.
However, the most recent analyses \cite{FOPR,Barger:2001nu,Abel:2001mc} using a generic set of 
flavor-diagonal phases and the three
competitive EDM bounds suggests that such cancellations are difficult to achieve if at all.
\item {\bf decoupling}: One can clearly reduce the constraints by lifting the scale of the
superpartners \cite{Nath:dn}. If one only considers the leading one-loop contributions to the EDMs, this 
can be done in ways that do not significantly affect the level of tuning 
in the Higgs sector, i.e. by keeping the 3rd generation scalars light, as is also required by
cosmological constraints \cite{fos}. However, various subleading effects then come into play and limit 
the extent to which the EDMs can be suppressed. One of our goals here will be to map out precisely how 
far EDMs can reach in terms of the superpartner scale.
\end{itemize}

Taken as a whole, one may observe that these scenarios could allow for various partial
suppressions of the EDMs but, although various models exists, within the generic supersymmetric framework
we currently lack a compelling symmetry argument for why the EDMs should be suppressed by 
many orders of magnitude as they are in the Standard Model. Thus, with the current level of experimental 
progress, we can expect many of these scenarios to be put to the test in the coming few years, 
providing ample motivation for further theoretical consideration of the EDM sensitivities. 

To summarize the current situation, recall that on restricting attention to the framework of the constrained
MSSM~\cite{cmssmus,cmssmthem}, one reduces the number of 
new fundamental phases to two, which can be chosen to be the phases of the $A$ and $\mu$ parameters at the
GUT scale. It is well known that the phase of $\mu$ is 
particularly effective in inducing EDMs, and therefore for reasonably generic sparticle spectra 
is tightly constrained, while the phase of $A$ can in principle be maximal (See {\em e.g.} \cite{WWII}). 
For example, this may be explained by the $\tan\beta$-enhancement of the one-loop EDMs 
induced by the phase of $\mu$, while the corresponding enhancement of the $A$-phase 
arises only at the two-loop level. In fact, as we shall discuss below, this argument is far from
watertight as other contributions can also be important.

The present work has three main goals. The first, and more technical, goal is to update the calculation 
of the EDMs 
by including certain two-loop corrections in the running of the soft-breaking parameters 
from the GUT to the weak scale. At first sight, the inclusion of the two-loop 
running should not substantially change the prediction for EDMs in terms of the SUSY phases. 
However, a complex $A$ parameter induces imaginary corrections to the gaugino masses,
which then induce the EDMs via one-loop SUSY threshold diagrams that are
enhanced by $\tan\beta$. This significantly modifies the EDM predictions as 
a function of the $A$-phase at the GUT scale and remarkably enough, in the case of 
even moderate $\tan\beta$, these 
new contributions can provide the dominate source for the electron EDM.

The second goal of the paper is to present an updated and complete analysis of
the constraints imposed by EDMs on the parameter space of the constrained MSSM (CMSSM). 
We focus on the CMSSM in order to incorporate full phenomenological analyses. However, 
for various reasons it seems that relaxing the universality assumption does not in fact 
significantly alter the conclusions \cite{Barger:2001nu,Abel:2001mc}, although this issue 
perhaps deserves further attention. 
We will update the corresponding results of \cite{Barger:2001nu,Abel:2001mc}, by incorporating the new 
Tl-EDM bound and most significantly by treating the qualitatively distinct large $\tan\beta$ regime, 
utilizing the results of \cite{dlopr}. 
The third and final goal is to emphasize the significant improvements in sensitivity, by orders of 
magnitude in
most cases, that are likely within a few years due to the current and next-generation EDM searches. 

The paper is organized as follows. In the next section we detail the new 
contributions to the EDMs induced primarily by two-loop running of the soft-breaking 
parameters. In Section 3, for completeness we give a short summary of the EDM formulae 
that relate the observables to the Wilson coefficients of the $CP$-odd effective Lagrangian, which are 
in turn determined by SUSY diagrams. We also outline the anticipated sensitivity 
of the next generation EDM experiments. Section 4 describes the numerical analysis 
within the CMSSM framework, and presents the limits on $CP$ violating phases as well as
the potential reach of EDMs in terms of the SUSY mass spectrum. Some concluding remarks
appear in section 5.

\section{RG evolution of the $CP$-odd invariants at two loops}

We begin by exhibiting a simplified formula for the one-loop contributions to the 
electron EDM in terms of the 
$CP$-violating SUSY phases \cite{FOPR},
\begin{eqnarray}
d_e = \fr{em_e}{16 \pi^2 M_{\rm SUSY}^2}\left[\left(\fr{5g_2^2}{24}
+ \fr{g_1^2}{24}\right) \tan\beta\sin\!\left[{\rm Arg} (\mu M_2 m_{12}^{2*})\right] 
+\fr{g_1^2}{12}\sin\!\left[{\rm Arg} (M_1^* A_e)\right]\right]\!,
 \;\;\;\;
\label{simplified}
\end{eqnarray}
where $M_1(M_2)$ are the U(1)(SU(2)) gaugino masses, 
all the couplings and masses are normalized at the weak scale, and 
we chose a SUSY parameter point with the absolute value of all 
soft breaking parameters in the selectron and gaugino sector equal to $M_{\rm SUSY}$.
This leads to a transparent overall normalization for $d_e$. The soft breaking parameter in
the Higgs sector, $m_{12}^2$, enters Eq.~(\ref{simplified}) via the relative phase
of the two Higgs vacuum expectation values. It is possible to choose 
$m_{12}^2$ (also denoted $B\mu$ or $b$ in an alternative notation) and consequently $v_1$ and 
$v_2$ to be real. 
It is easy to see why the phases of $\mu$ and the Wino mass 
give larger contributions to $d_e$ than does the phase of $A$. For example,
\be
\fr{d_e(\th_\mu,\th_{M_2})}{d_e(\th_{A_e},\th_{M_1})} \simeq \fr{5 g_2^2}{2 g_1^2} \tan\beta
\times \fr{\sin(\th_\mu +\th_{M_2})}{\sin(\th_{A_e} - 
\th_{M_1})}\simeq 10 \tan\beta \times\fr{\sin(\th_\mu +\th_{M_2})}{\sin(\th_{A_e} - \th_{M_1})},
\label{enhancement}
\ee
which implies that the sensitivity to the phase of $\mu$ is enhanced by two orders of magnitude  
relative to the phase of $A$, even for moderate values of $\tan\beta$. For hadronic EDMs this 
effect is also present, although the advantage of $\th_\mu$ is less pronounced 
because both phases can generate EDMs proportional to $g_3^2$. The enhancement 
(\ref{enhancement}) provides the primary explanation for why the phase of $\mu$ is 
more severely constrained than the phases of the $A$-parameters.

Even with the restrictions of flavor universality and proportionality, the number of 
independent $CP$-odd phases can be rather large. Working within the CMSSM scenario, 
one assumes a common phase for the trilinear parameters at the GUT scale, and a common phase 
for the gaugino masses. This reduces the number of physical $CP$-odd phases to two
and a traditional choice of basis at the GUT scale comprises $\theta_\mu$ and $\theta_A$:
\begin{eqnarray}
 && ~~~~~~~~~~~(\th_A \equiv {\rm Arg} (A_f), \th_\mu \equiv {\rm Arg}(\mu)); \\
 && ~~~~~~~~~~~~~~~~~~~~~{\rm Arg} (M_i) = 0; \nonumber\\
 && \th_{12}\equiv{\rm Arg} (m_{12}^2)~~\mbox{-- {\rm tuned to ensure}}
 ~{\rm Arg}(m_{12}^2)_{EW}=0.
\label{choice}
\end{eqnarray}
The utility of this basis lies in the invariance of the ${\rm Arg} (M_i) = 0$ 
condition under one-loop renormalization group evolution, and also the  
invariance of $\theta_\mu$. If needed, the basis (\ref{choice}) 
can be rotated to any other convenient basis in the CMSSM framework by  
appropriate phase redefinitions of the matter and gauge superfields. 
The choice of Arg$(m_{12}^2)$ at the GUT scale, which is {\em not} invariant under 
RG evolution \cite{Garisto:1996dj,fo2}, is such as to ensure the reality of the 
Higgs vevs. Thus, choosing different values for $\th_\mu$ and $\th_A$ requires a re-adjustment 
of $\th_{12}$ at the GUT scale in these conventions which,  although aesthetically unappealing, 
simplifies calculations by eliminating the dependence on $\th_{12}$ of the EDMs. 

\begin{figure}
\centerline{\includegraphics[width=7cm]{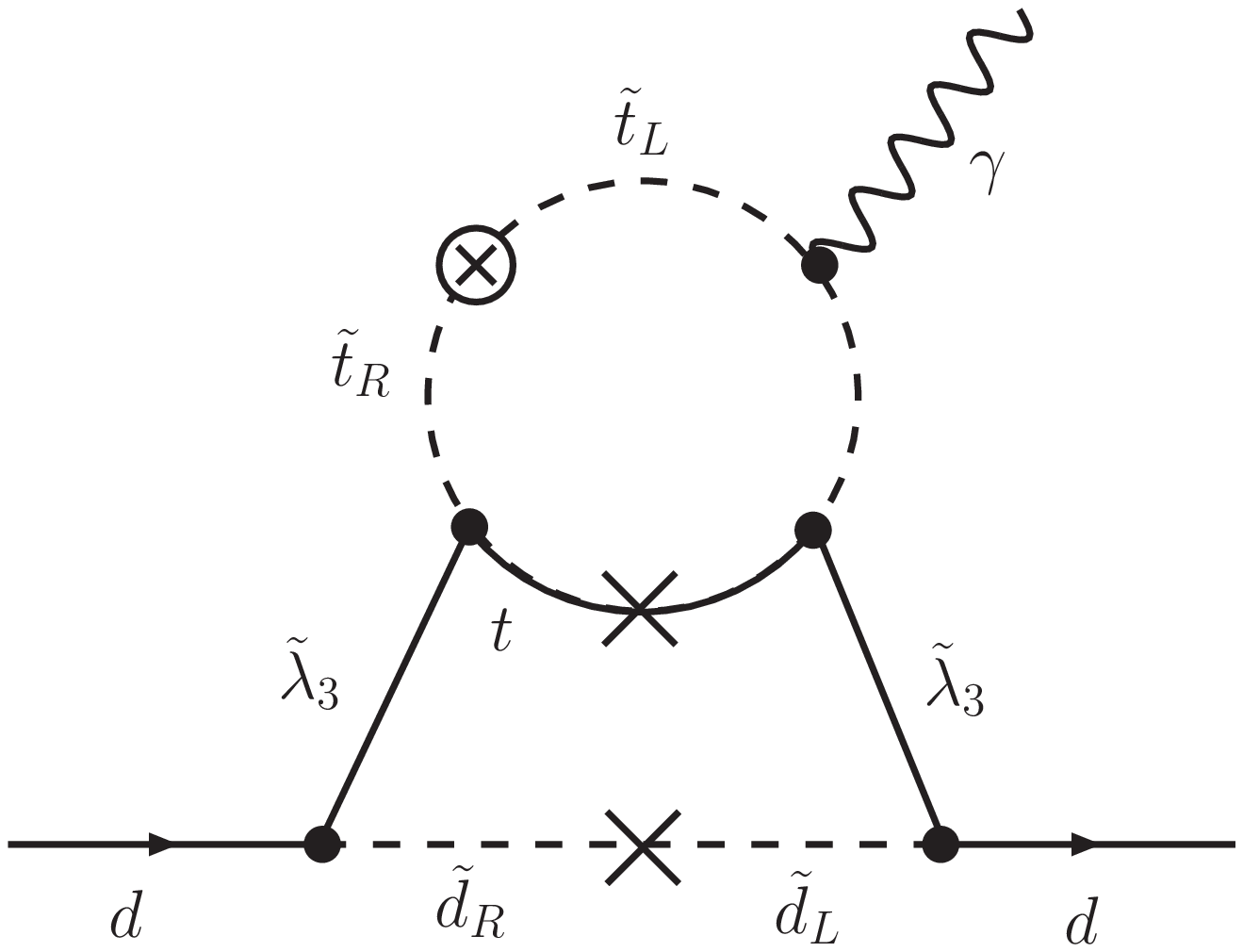}\includegraphics[width=7cm]{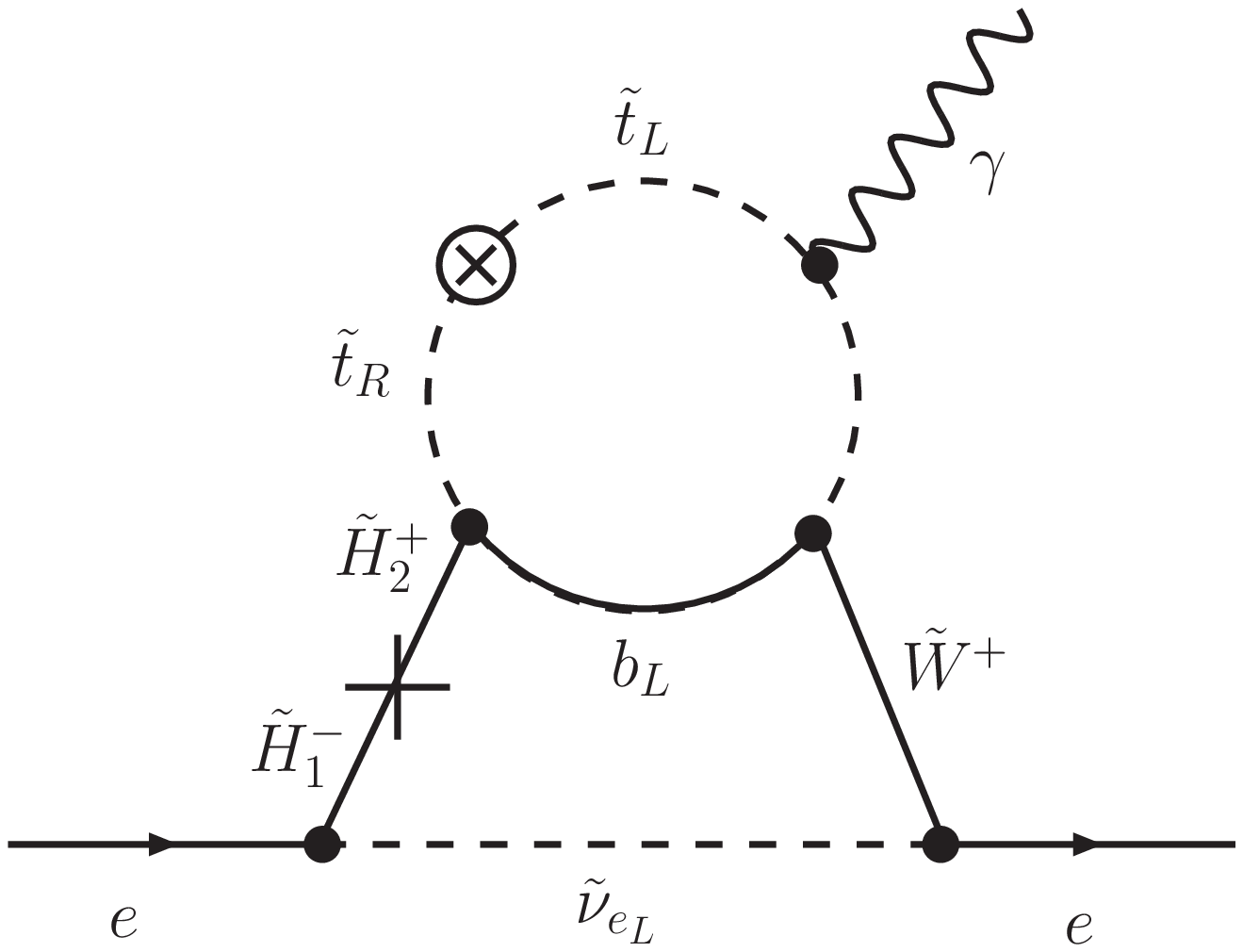}}
 \caption{\footnotesize On the left is an example of a $\tan\beta$-enhanced two-loop threshold correction 
to $d_d(\th_A)$. If $\mu$ is real, the phase enters via the circled left-right stop insertion. On the right we 
show a more significant $\tan\beta$-enhanced threshold correction to $d_e(\th_A)$ \cite{Pil99b}.}
\label{f1} 
\end{figure}


\subsection{Further contributions to $d_e(\th_A)$}

With this basis in mind, consider the relative enhancement of the 
$(\th_\mu+\th_{M_2})$ contribution to $d_e$ as apparent in (\ref{enhancement}). One observes that for moderate
values of $\tan\beta$ it is sufficiently large to counteract a further loop
suppression factor. Thus, in the absence of a $\mu$-phase, one may enquire whether higher-loop corrections, 
induced by $\th_A$, may become sizable in this way at large $\tan\beta$ and compete with the standard
one-loop contribution to $d_e(\th_A)$. A number of 2-loop threshold corrections to the fermion EDMs 
have been considered in the MSSM literature, notably supersymmetric variants of the Barr-Zee 
diagrams \cite{ckp} which involve the Higgs sector. However, a number of other corrections of this 
type exist. Indeed, its clear that top-stop threshold corrections on the gaugino line, as 
shown for $d_d$ in Fig.~\ref{f1}a, may give a $\tan\beta$--enhanced contribution to
fermion EDMs $d_f(\th_{A_t})$ and quark color EDMs (CEDMs) $\tilde{d}_q(\th_{A_t})$. However, its 
relative size compared to the conventional one-loop term, ignoring hierarchies in $\mu$ and $A$, 
is of order $\tan\beta(m_t/m_{\tilde{t}})^2/(16\pi^2)$, and thus not particularly sizable. 

These corrections would clearly be much more important in the case of $d_e$ were it possible for a dependence on
$\th_A$ to manifest itself via the chargino diagram, and make full use of the enhancement 
factor of ${\cal O}(10\tan\beta)$ in (\ref{enhancement}). Threshold corrections of this type were noted
by Pilaftsis \cite{Pil99b}, and the most prominent example is shown in Fig.\ref{f1}b. The internal stop-bottom
loop provides a complex wavefunction correction of order $\Si \sim m_t A_t^*/(16\pi^2 m_{\tilde{t}}^2)$ 
which mixes $\tilde{H}_2$ with $\tilde{W}^+$. Taking the limit where the stops are heavy, the resulting 
contribution to $d_e$ scales as $\De d_e \sim m_e\tan\beta {\rm Im}(A_t)/((16\pi^2 m_{\tilde{t}})^2\mu)$, which
despite the loop supression can be comparable with, or even larger than, the 1-loop neutralino diagram. 
Moreover, such contributions are often considerably larger than the more widely known Barr-Zee type diagrams, 
and deserve to be included in more general analyses. However, in the restricted framework of the CMSSM, while
competetive with other contributions they are somewhat suppressed by RG effects which generically render the 
ratio Im$(A_t)m_t/m_{\tilde{t}}^2$ rather small at the weak scale. 
Indeed, we will not need to study them in detail because, remarkably enough, it turns out 
that there is an even larger class of contributions which are less sensitive to detailed features of the spectrum. 

In particular, one can consider a diagram where in the top-stop
loop, e.g. of Fig.\ref{f1}a, we tie the Higgs lines together to generate a two-loop correction to the 
running of the gluino mass. While further loop suppressed, this diagram is enhanced by a large
RG log and the relative contribution of such a diagram compared to Fig.\ref{f1}a is 
of order $\ln(\mu_{\rm GUT}/M_Z)(m_{\tilde{t}}/m_t)^2/(16\pi^2)$, which can easily be greater than
one. In the case of quark EDMs, such corrections are still generally negligable. However, when such
RG corrections are accounted for on the Wino line of the chargino diagram, the corrections to $d_e(\th_A)$
as shown in Fig.\ref{f2} can be sizable. Despite the large loop supression, it benefits from the 
enhancement factor in (\ref{enhancement}), a large RG log, and since the insertion is divergent it is
not suppressed by the generically large squark masses. Remarkably, within the CMSSM and for moderate to large
$\tan\beta$, it seems that this 3-loop effect can easily be larger than the leading 1-loop threshold 
correction to $d_e(\th_A)$, and also the 2-loop threshold corrections in Fig.\ref{f1}b. Thus 
it will have important consequences for the EDMs, as we will now discuss more quantitatively below.

\subsection{Two-loop RG evolution and EDMs}

To be more precise, at two-loop order, imaginary corrections to the gaugino masses may be generated
if Im$A$ is nonzero. The corresponding part of the RG equation for $M_i$ \cite{mv},
\be
 \frac{d}{dt} {\rm Im}(M_i) = \frac{2 g_i^2}{(16\pi^2)^2}\sum_{f=u,d,e} C^f_i {\rm Im}(A_f)\,{\rm  Tr} 
  ({\rm \bf Y}_f^\dagger {\rm \bf Y}_f),
\ee 
with (in the $\overline{\rm DR}$ scheme)
\be
 C_i^f = \left(\begin{array}{ccc}
                 26/5 & 14/5 & 18/5 \\
                 6 & 6 & 2 \\
                 4 & 4 & 0 
               \end{array}\right),
\ee
spans all the flavors, but is dominated by the top, and for large $\tb$ bottom and 
tau, contributions.

\begin{figure}
\centerline{\includegraphics[width=14cm]{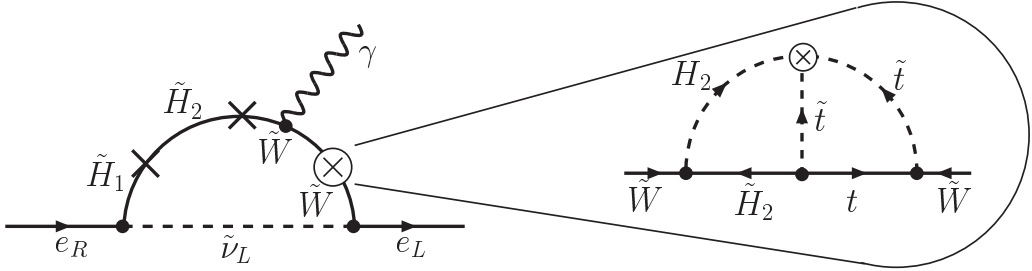}}
 \caption{\footnotesize A two-loop RG correction to Im$M_2$, induced by $\th_A$ which contributes
to $d_e(\th_A)$ via the chargino threshold diagram (shown in the mass-insertion approximation). 
Despite the loop suppression, the result is
enhanced by $\tan\beta$, a large RG log, and a ratio of couplings, and thus can be larger
than the one-loop neutralino contribution as discussed in the text.}
\label{f2} 
\end{figure}
 
While we will not attempt a full treatment of the two-loop RG equations, we will 
take into account the relevant contributions that 
generate Im$(M_i)$ at the weak scale. In order to estimate the order of magnitude of this effect, 
we can consider a simple scenario where we focus on the evolution of $M_2$, and retain only the 
largest (top) Yukawa term. One then finds, $(d/dt) {\rm Im}(M_2) \simeq 3Y_t^2 \al_2/(16\pi^3) {\rm Im}(A)$, 
which to leading-log order integrates to,
\be
 {\rm Im}(M_2(M_Z)) \simeq -\frac{3\al_2(\mu_{\rm 
GUT})}{32\pi^3} {\rm Im}(A_0) 
    \ln\left(\frac{\mu^2_{\rm GUT}}{M_Z^2}\right) \ \sim \ -6 \times 10^{-3} {\rm Im}(A_0),
    \label{estimate}
\ee
where we have inserted the GUT-scale value for $\al_2$, but used the weak scale value for $Y_t$. This
estimate, which accounts only for the top-stop contribution, provides a reasonably accurate 
approximation for most regimes. The two-loop suppression factor indicates why resummation
of the large logs does not change the result significantly.
Written in basis-independent form, the estimate (\ref{estimate}) expresses corrections 
to three weak scale $CP$-odd invariants, Arg$(\mu M_i m_{12}^{2*})$, as a function of the 
GUT scale invariants Arg$(M_{i}A)$.

As noted above, a particular example of the relevant contributions to the EDMs is shown explicitly in
Fig.~\ref{f2}. It combines the two-loop running of the soft parameters with the one-loop SUSY threshold
contribution to the EDM. It is clear that to a certain degree the effect of Im$(M_i)$ mimics
Im$(\mu)$ inside such a diagram, and thus is most prominent when
$\vec{\theta}_{\rm GUT} \equiv (\th_\mu,\th_A) = (0,\th_A)$. These contributions are clearly also
present for quark EDMs and CEDMs, but as discussed above are less competitive as the corresponding
enhancement factor is an order of magnitude smaller than (\ref{enhancement}).

The result (\ref{estimate}) has a significant effect on the calculation of EDMs induced by 
$\th_A$ for several reasons:
\begin{itemize}
\item The two-loop running destroys the universality of the gaugino mass phases at the weak scale, 
which therefore have to be taken into account in EDM calculations. 
\item The resulting size of the imaginary parts of $M_i$ appears to be small, 
$O(10^{-2}-10^{-3})\times M_{\rm SUSY}$ (\ref{estimate}). However, in combination with the 
enhancement factor $\sim 10 \tan\beta$ (\ref{enhancement}), the resulting correction to
$d_e$ can be $O(1-10)$, for moderate-to-large $\tb$, as compared to the ``standard'' 
result for $d_e(\th_A)$. 
\item The two-loop  RG induced correction to $d_e(\th_A)$ has the opposite sign relative to 
the standard result, providing a regime where the two contributions can cancel each other. 
\end{itemize}

Putting together the various numerical factors, from RG-running and the SUSY threshold-induced 
EDM, we see that the new contributions induced by Im$(M_i(\th_A))$ correspond to 
a three-loop correction which is enhanced by ${\cal O}(1-10)\tb \times \ln(\mu^2_{\rm GUT}/M_Z^2)$,
and as emphasized above is often larger than the leading 1-loop threshold correction and
various additional 2-loop threshold corrections.
The question then arises as to whether there are other three-loop effects with a comparable 
enhancement, {\em i.e.}
$\propto \tan^2\beta$. It is well known, however, that such contributions do not 
exist. Enhancement by a single power of $\tb$ results from $d_e \propto y_e \langle H_2\rangle 
\propto m_e\tan\beta$. Any additional power of $\tb$ can arise only from
additional insertions of the Yukawa couplings of charged leptons or down-type 
quarks, which cannot lead to a numerical enhancement. 

\section{An EDM summary}

In this section, we briefly recall the dependence of the observable EDMs on the most
significant flavor-diagonal $CP$-odd operators at 1 GeV (see \cite{PRrev} for a recent review). 
Up to dimension six, the
corresponding effective Lagrangian takes the form,
\begin{eqnarray}
\label{leff}
{\cal{L}}_{eff} &=& \frac{g_s^2}{32\pi^{2}}\ \bar\theta\
G^{a}_{\mu\nu} \widetilde{G}^{\mu\nu , a} \nonumber\\  && -
\frac{i}{2} \sum_{i=e,u,d,s} d_i\ \overline{\psi}_i (F\sigma)\gamma_5 \psi_i  -
\frac{i}{2} \sum_{i=u,d,s} 
\widetilde{d}_i\ \overline{\psi}_i g_s (G\sigma)\gamma_5\psi_i \nonumber\\
  && + \frac{1}{3} w\  f^{a b c} G^{a}_{\mu\nu} \widetilde{G}^{\nu \beta , b}
G^{~~ \mu , c}_{\beta}+ \sum_{i,j} C_{ij} (\bar{\ps}_i \ps_i) (\ps_j i\ga_5 \ps_j)+ \cdots
\end{eqnarray}
For the four-fermion operators, the sum runs over all light fermions, and one could 
in principle include additional tensor structures \cite{KKY,KK2,CPfourferm,PH}. In practice, 
since these operators require a double helicity flip, implying additional $(m_q/v_{EW})^2$ suppression 
factors, they are generally negligible. The exception to this rule is at large $\tan\beta$, where
$C_{ij}$ grow as $(\tan\beta)^3$, and they can compete with the other sources in certain
parameter regimes \cite{LP,dlopr}.

In supersymmetric models with $CP$-violating phases larger than 
$\sim 10^{-8}$ in the flavor-diagonal 
sector the only reasonable strategy to avoid the strong 
$CP$ problem is to postulate the existence of the Peccei-Quinn 
relaxation mechanism \cite{PQ}, which removes the theta term from 
the list of contributing operators. We will adopt this strategy here.

We will now turn to a quick synopsis of the dependence of the observable EDMs on the 
remaining dimension five and six operators. The physical observables can be conveniently
separated into three main categories, depending on the physical mechanisms via which an
EDM can be generated: EDMs of paramagnetic 
atoms and molecules, EDMs of diamagnetic atoms, and the neutron EDM.

\subsection{EDMs of paramagnetic atoms -- thallium EDM}

Among various paramagnetic systems, the EDM of 
the thallium atom currently provides the best constraints. Atomic calculations
\cite{LiuKelly,MP1,MP2} (see also Ref. \cite{KL} for a more complete list) link the
atomic EDM with $d_e$ and various $CP$-odd electron-nucleon interactions, of which
we shall only consider the most relevant, namely $C_S\bar e i \gamma_5 e \bar NN$,
\be 
d_{\rm Tl} = -585 d_e -   e\ 43 ~{\rm GeV} C_S^{sing},
\label{dtl}
\ee
where we have furthermore retained only the leading isosinglet component of $C_S$.
The relevant atomic matrix elements are calculated to within 
$10-20\%$ accuracy \cite{fgreview}.\footnote{It is worth emphasizing that, due to
the contributions from $C_S$, a bound on more than one paramagnetic species is required
to infer a direct bound on $d_e$. Thus, a conservative and model-independent bound on 
$d_e$ cannot be derived from the bound on $d_{\rm Tl}$ alone, and would be $\sim$ two orders of
magnitude weaker than the commonly quoted value obtained by dropping the dependence on
$C_S$ in (\ref{dtl}).}

The dependence of $C_S$ on the four fermion sources $C_{ie}$, for $i=d,s,b$, 
can be expressed in the form \cite{LP,dlopr},
\be
C_S^{sing} = C_{de}\frac{29~{\rm MeV}}{m_d}+ C_{se}\frac{\kappa \times 220~{\rm MeV}  }{m_s}+ 
C_{be}\frac{66~{\rm MeV}(1-0.25 \kappa )}{m_b},
\ee
where $\kappa \equiv \langle N|m_{s}\overline{s}s|N\rangle /220$ MeV, and we will adopt the
value obtained using NLO $\ch$PT, $\ka \sim 0.50 \pm 0.25$ \cite{KBM}. Larger values 
for the pion-nucleon sigma term $\si_{\pi N}$, advocated recently (see e.g. \cite{sigma_large}), 
would result 
in a somewhat larger value for $\kappa\sim 1.55$.

\subsection{Neutron EDM}

The calculation of the neutron EDM in terms of the Wilson coefficients of 
Eq.~(\ref{leff}) represents a difficult non-perturbative problem, and thus the precision 
currently obtainable is not competitive with the results outlined above for $d_{\rm Tl}$. 
Nonetheless, $d_n$ plays a crucial role in constraining $CP$-odd sources in the quark sector,
and we will make use of the results obtained using QCD sum rule techniques \cite{PR,DPR} (see
\cite{CDVW,PdR,FOPR,hs} for alternative approaches using chiral techniques), that we briefly 
recall below.

In the presence of the PQ mechanism, it is natural to expect that the dominant contribution 
to the neutron EDM comes from the EDMs and CEDMs of the light quarks. Within the 
sum-rules framework, PQ relaxation also suppresses the sea-quark contribution at leading 
order, and leads to the following result \cite{PR}:
\be
 d_{n}(d_q, \tilde d_q) = (1.4 \pm 0.6)(d_d-0.25d_u) + (1.1 \pm 0.5)e(\tilde d_d + 0.5\tilde d_u).
\label{dn1}
\ee
The quark vacuum condensate, $\langle \bar qq\rangle = (225 \, {\rm MeV})^3$, 
has been used in this relation -- the proportionality to $d_q\langle \bar qq\rangle \sim 
 m_q\langle \bar qq\rangle \sim f_\pi^2m_\pi^2$ removes 
any sensitivity to the poorly known absolute value of the light quark masses. 
Here $\tilde d_q$ and $d_q$ are to be normalized at the hadronic scale 
which we assume  to be 1 GeV. 
Note also that the quark masses used for the SUSY calculations of $d_q$ and $\tilde d_q $ 
should be taken at the weak scale, where their numerical values are  smaller 
than the low energy values by a factor of $\sim 0.35$, {\em e.g.} 
$ m_d(M_Z) \simeq 9.5\,{\rm MeV} \times 0.35$. 

The dimension six sources, $w$ and $C_{q_1q_2}$, also contribute to the neutron EDM but are more
problematic to handle within the sum-rules framework. One can nonetheless obtain estimates
that are useful for assessing the regimes in which the contributions in (\ref{dn1}) are
dominant. One finds \cite{DPR},
\be
  d_n(w) \sim 20\,{\rm MeV}\times e~w.
\label{dn2}
\ee
This estimate is assessed to be valid within a factor of 
2--3 \cite{DPR}. When $\tan\beta$ is large, the dominant contributions to $w$ arise via
threshold corrections from the CEDMs of heavy quarks \cite{CKKY,ALN}, e.g. 
$\de w(\tilde{d}_b) \simeq g_s^3\tilde d_b(m_b)/(32 \pi^2 m_b)$.

In general there are also many possible contributions from four-fermion sources. However,
since these contribute only at large $\tan\beta$, the most relevant are $C_{bd}\simeq -C_{db}$ and 
$C_{sb}\simeq -C_{bs}$. In this case, one can obtain an estimate of the form \cite{dlopr},
\be
 d_n(C_{bd})\sim e~0.7\times 10^{-3} {\rm GeV}^2\,\frac{C_{bd}}{m_b},
\label{dn3}
\ee
which again is primarily useful as a means to estimate regimes where large corrections
to (\ref{dn1}) are possible.

\subsection{EDMs of diamagnetic atoms -- mercury EDM}

EDMs of diamagnetic atoms, i.e. atoms with total 
electron angular momentum equal to zero, also provide an important test of $CP$ 
violation \cite{KL}. The current limit on the EDM of mercury 
\cite{Hg} furnishes one of the most sensitive constraints on SUSY 
$CP$-violating phases \cite{FOPR}. However, the calculation of $d_{\rm Hg}$ is 
undoubtedly the most difficult as it requires QCD, nuclear, and also atomic input. 

The atomic EDM of mercury arises from several important sources, namely, 
the Schiff moment $S$ \cite{Schiff}, the electron EDM $d_e$, and also 
electron-nucleon interactions (see, e.g. Ref. \cite{KL}
for a comprehensive review). Schematically, the mercury EDM can 
be represented as
\be
d_{\rm Hg} = d_{\rm Hg}\left( S[\bar g_{\pi NN}(\tilde d_i, C_{q_1 q_2})], \, 
C_S[C_{qe}],\, C_P[C_{eq}],\, d_e \right),
\label{dhgsch}
\ee
where $\bar{g}_{\pi NN}$ collectively denotes the 
$CP$-odd pion-nucleon couplings, and $C_S$ and $C_P$ denote respectively the 
couplings $\bar e i \gamma_5 e \bar NN$ and $\bar e e \bar N i\ga_5N$.
The Weinberg operator does not provide 
any appreciable contribution to $d_{\rm Hg}$ because its 
contribution to $\bar{g}_{\pi NN}$
is suppressed by an additional chiral factor  of 
$m_q/1\,{\rm GeV}\sim 10^{-2}$.

A number of the atomic and nuclear components of the calculation have been updated recently,
specifically $d_{\rm Hg}( S)$ \cite{Hgnew} and $S(\bar g_{\pi NN})$ \cite{DS}. We will also
make use of an updated result for $\bar{g}_{\pi NN}(\tilde d_i)$ obtained by combining chiral
and sum-rules techniques \cite{P}, 
\be
 \bar{g}_{\pi NN}^{(1)}(\tilde{d}_q)  \sim 4^{+8}_{-2} (\tilde{d}_u - \tilde{d}_d) [{\rm GeV}^{-1}],
\ee
where the relatively large overall uncertainty arises from significant cancellations between
the leading contributions. Combining this result with the contribution of four-fermion
operators \cite{dlopr}, and the atomic and nuclear parts of the calculation, we obtain
\begin{eqnarray}
d_{\rm Hg} &=& 7\times 10^{-3}\,e\,(\tilde d_u - \tilde d_d)  + 10^{-2}\, d_e
 \nonumber \\
\label{Hgmaster}
 && -1.4\times 10^{-5}\,e\,{\rm GeV}^2 
\left( \fr{0.5 C_{dd}}{m_d} +3.3 \kappa \fr{ C_{sd}}{m_s}+
\fr{C_{bd}}{m_b}(1-0.25\kappa)\right)
\\\nonumber
 && +3.5\times 10^{-3}{\rm GeV}\, e\, C_S \;+\;
4 \times  10^{-4}{\rm GeV}\, e\, C_P.
\end{eqnarray}
The contributions in the second and third lines are significant in the MSSM
only for large $\tan\beta$.

The most valuable feature of
$d_{\rm Hg}$ is its sensitivity to the triplet combination of
CEDM operators $\tilde d_i$, which surpasses the sensitivity of the neutron EDM 
to this combination by a factor of a few. Moreover, despite the fact that the overall coefficient
is only known with relatively poor precision, the
dominant dependence on the $(\tilde d_u - \tilde d_d)$ combination
ensures that these uncertainties enter as an overall factor
and therefore do not significantly alter the 
shape  of the unconstrained band of the parameter space in
the $\theta_\mu - \theta_A$ plane.

\subsection{Future experimental sensitivity}

The experimental situation is currently very active, and a number of new EDM experiments are
already in development which promise to significantly improve the level of sensitivity by several
orders of magnitude in the next few years. As part of our analysis in the next section, we will present 
the reach of these next-generation experiments within the CMSSM, given the following assumed sensitivity
in the three classes of EDMs on which we have been focussing:
\ba
 |\Delta d_{\rm e}| &<& 3 \times 10^{-29} e\, {\rm cm} \nonumber\\
|\Delta d_{\rm D}| &<& 2   \times 10^{-27}  e\, {\rm cm}   \nonumber  \\
|\Delta d_n|  &<&   1\times 10^{-27} e\, {\rm cm}.
\label{fexplimit}
\ea
These sensitivities were chosen conservatively according to expectations for the first physics runs
of the relevant experiments discussed below. It is possible that the ultimate sensitivity will be
be at least an order of magnitude better in most cases.

The first constraint shown in (\ref{fexplimit}) corresponds to a relatively conservative estimate of 
the sensitivity achievable within three active projects utilizing respectively two polarizable 
paramagnetic molecules, PbO \cite{PbO} and YbF \cite{YbF}, and also a solid state 
system \cite{Lamoreaux}. The first two of these primarily obtain their enhanced sensitivity 
from the fact that the internal electric field, to which the electron EDM is sensitive, is 
further enhanced by the polarization of the molecule. One should bear in mind that, as for Tl,
these experiments do not directly bound $d_e$ and electron-nucleon operators may also be relevant.
PbO and YbF are polarized by the external field in a nonlinear way and thus do not have EDMs
as such. However, the induced shift in the Larmor precession frequency can still be written in the
form,
\be
 \hbar \De \om_{\rm L} \sim {\cal E}_{\rm eff} d_e  + {\cal O}(C_S).
\ee
We are not aware of any
calculations for the $C_S$-dependence in this case, while the quoted sensitivity in (\ref{fexplimit})
comes from ignoring this term and using estimates for the effective field ${\cal E}_{\rm eff}$
\cite{kozlov} and the sensitivity to $\om_{\rm L}$ \cite{PbO,YbF}. As alluded to above, if 
no further systematic issues arise, 
the final sensitivity could in fact be several orders of magnitude better than this quoted 
result \cite{PbO,Lamoreaux}. Given that the relevant atomic calculations are only partially
completed, we will simply scale up the result for Tl and restrict ourselves to a relatively
low value of $\tan\beta$, in order to ameliorate the problem of unknown corrections from $C_S$. 
Nonetheless, one should
be aware that the molecular matrix element relating $C_S$ and $\omega_{\rm L}$, although enhanced 
by the same factor $Z^3$ will not be a simple rescaling of the Tl result, 
and thus the final dependency $\omega_{\rm L}(d_e,C_S)$ may be somewhat 
different from $d_{\rm Tl}(d_e, C_S)$. 
Clearly further progress on atomic calculations for this system would be welcome.

The second constraint in (\ref{fexplimit}), refers to the proposed search for the deuteron EDM using
the BNL storage ring \cite{bnl}. This proposal seems our best chance for significant improvement in the 
diamagnetic sector. The deuteron is not a diamagnetic system of course, but the EDM turns out
to be primarily sensitive to $\bar{g}^{(1)}_{\pi NN}$ (assuming $\bar\th$ is removed by PQ rotation)
\cite{kk,lopr,ltk},
\be
 d_D \simeq d_n + d_p -(1.3 \pm 0.3) e  \bar{g}^{(1)}_{\pi NN}(\tilde{d}_q) [{\rm GeV}^{-1}] 
\ee
and thus it plays a similar role in terms of
its sensitivity to underlying $CP$-odd sources. Indeed, it has the significant advantage
in this regard of not requiring complicated many-body nuclear calculations.

The final constraint in (\ref{fexplimit}) refers to the expected (initial) sensitivity 
of several experiments currently in development, at LANSCE \cite{lansce},
at ILL \cite{ILL}, and at PSI \cite{PSI}, 
searching for an EDM of the neutron. The ultimate sensitivity could again be an order of magnitude better.

\section{Numerical Results and EDM constraints}

Before we turn to the numerical analysis of EDMs within the CMSSM framework, for completeness we
now list the relevant SUSY contributions to the operators in (\ref{leff}) that must be included:

\subsection{SUSY contributions to \boldmath{$CP$}-odd higher-dimensional operators}

Within the effective theory formulated at the weak scale, or rather the mass scale of the 
superpartners, the $CP$-odd operators in (\ref{leff}) can be induced by the following threshold corrections,
all of which we will include in our analysis.

\bigskip
{\it Fermion EDMs and CEDMs}
\begin{itemize}
\item {\bf 1-loop}: At one-loop order the electron EDM receives contributions from
$\tilde{\ch}^{-}-\tilde{\nu}_e$ and $\tilde{\ch}^0 - \tilde{e}^{-}$ loops (see e.g. \cite{IN}).
The EDMs and CEDMs of quarks receive analogous contributions, and in addition have generically
dominant contributions from gluino-squark loops (see e.g. \cite{IN}).
As we argued in Section 2, these one loop results {\em must} be formulated in the 
basis where Im$M_i$ at the weak scale are taken into account.

\item {\bf 2-loop}: At 2-loop order, non-negligible Barr-Zee-type Higgs-mediated graphs contribute
to EDMs and CEDMs \cite{ckp}. The induced photon couplings, e.g. $AFF$ and $HF\tilde{F}$, arise
through sfermion and chargino loops, with dominant contributions from stop (and, for large $\tan\beta$, 
sbottom and stau \cite{dlopr}) loops. These diagrams can be significant when the first generation
sfermions are heavy.\footnote{Note that additional 2-loop threshold corrections, such as 
Fig.\ref{f1}b \cite{Pil99b}, can in principle be larger than the Barr-Zee type contributions, 
and thus a full analysis of 2-loop threshold corrections appears warranted. However, within the CMSSM,
we have verified that the complex wavefunction-induced correction $\Si$ \cite{Pil99b} to the chargino mass matrix
is, even when non-negligible compared to other conventional sources, still generally subleading relative to 
complex 2-loop RG corrections to the 1-loop diagrams.}  
\end{itemize}

\bigskip
{\it Weinberg operator}
\begin{itemize}
\item {\bf 2-loop}: The Weinberg operator \cite{W} also contributes to
$d_n$ via two-loop stop-top-gluino contributions (see e.g. \cite{IN}), particularly in the 
regime where the first generation sfermions are heavy. As noted earlier, the Weinberg operator may
also receive non-negligible corrections from heavy quark CEDMs under RG evolution to low
scales \cite{CKKY,ALN,dlopr}.
\end{itemize}

\bigskip
{\it Four-fermion operators}
\begin{itemize}
\item {\bf 1-loop}: We account for the leading $\tan\beta$--enhanced contributions \cite{LP}, as computed
in \cite{dlopr}, induced by scalar or pseudoscalar Higgs exchange, with a $\tan\beta$-enhanced 
coupling of down-type fermions to $H_2$ induced by various 1-loop threshold corrections. There
are also similar Higgs exchange contributions with loop-induced $H-A$ mixing that are generically
sub-dominant.
\end{itemize}

These contributions are quite generic in any SUSY model of new physics at the TeV scale.
We turn now to a numerical analysis of the EDM sensitivity to phases and SUSY masses 
within the framework of the CMSSM. 

\subsection{EDM reach and constraints on $CP$-odd phases in CMSSM}

In this section we analyze all three  
observables, $d_{\rm Tl}$, $d_n$ and $d_{\rm Hg}$, 
within the CMSSM, defined by the following set of universal SUSY parameters at the GUT scale:
\begin{equation}
\left\{\tan\beta,~m_0 ,~ m_{1/2},~\vert A_0 \vert,~ \theta_A,~ 
  \theta_\mu\right\},
\end{equation}
where $m_{1/2}$ and $m_0$ are the GUT scale universal gaugino and scalar
masses respectively.
The magnitude of the $\mu$--parameter and the pseudoscalar mass
are determined by the radiative electroweak symmetry breaking conditions.
We will present the analysis separately for each of the two physical phases, 
$\th_\mu$ and $\th_A$, which are present under these assumptions.

Let us recall that RG running from the GUT scale in this scenario introduces 
considerable mass splittings in the spectrum of superpartners; for example, 
the gluino becomes much heavier than (roughly triple) the rest of the gauginos,
while the squarks also become quite heavy,  $m_{\rm sq}^2(M_Z) \simeq m_0^2 + 6 m_{1/2}^2 + O(M_Z^2)$.
Of particular importance here, as emphasized above, is that the physical phases also run,
and mix, under the RG. The two-loop RGE for the $\mu$ parameter is real, and therefore the phase of $\mu$
does not run; its low energy value is equal to its input value at the GUT scale.
On the other hand, the phases of all of the $A$ parameters do run and are
primarily affected by the 3rd generation Yukawa couplings. As discussed earlier,
although we are free to set the phases of the gaugino masses to zero
at the GUT scale, the 2-loop RGEs induce phases for $M_{1,2,3}$ when the phases of 
$A$ are not zero. The mixing of physical phases that this induces will be 
illustrated below.

In what follows, we present our numerical results for the EDMs in the context of the
CMSSM.

\subsubsection{Phase constraints in benchmark scenarios}

Detailed phenomenological studies have found it useful to concentrate
on very specific CMSSM parameter choices.  To this end, several
sets of benchmark points have been established  \cite{oldbench,SPS,newbench}.
A set of proposed post-LEP benchmark
scenarios \cite{oldbench} were chosen to span the CMSSM parameter space
in regions where all phenomenological constraints are satisfied.  These include
obtaining a cosmological relic density with the limits established by WMAP~\cite{wmap}.
Several points were chosen to lie in the
`bulk' region at small $m_{1/2}$ and $m_0$, while others  are spread along
either the $\chi-{\tilde \tau}$ coannihilation `tail' at larger $m_{1/2}$ 
or in rapid-annihilation `funnels' where the relic
density is controlled by $s$-channel annihilation of the LSP through the heavy
Higgs scalar and pseudo-scalar at large $m_{1/2}$ and $m_0$ at large $\tan \beta$. 

\begin{figure}
\centerline{\includegraphics[width=7cm]{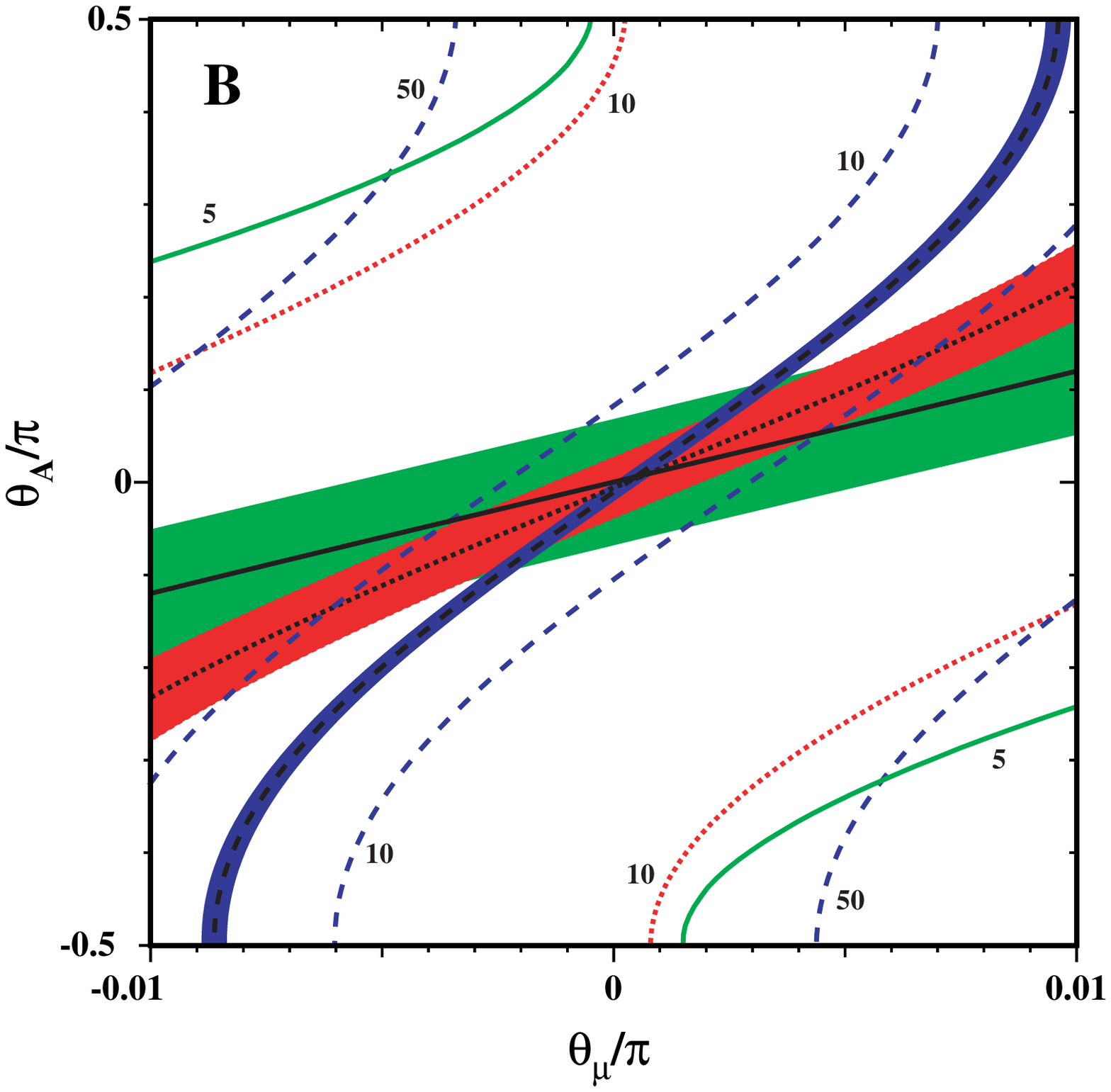}
\includegraphics[width=7cm]{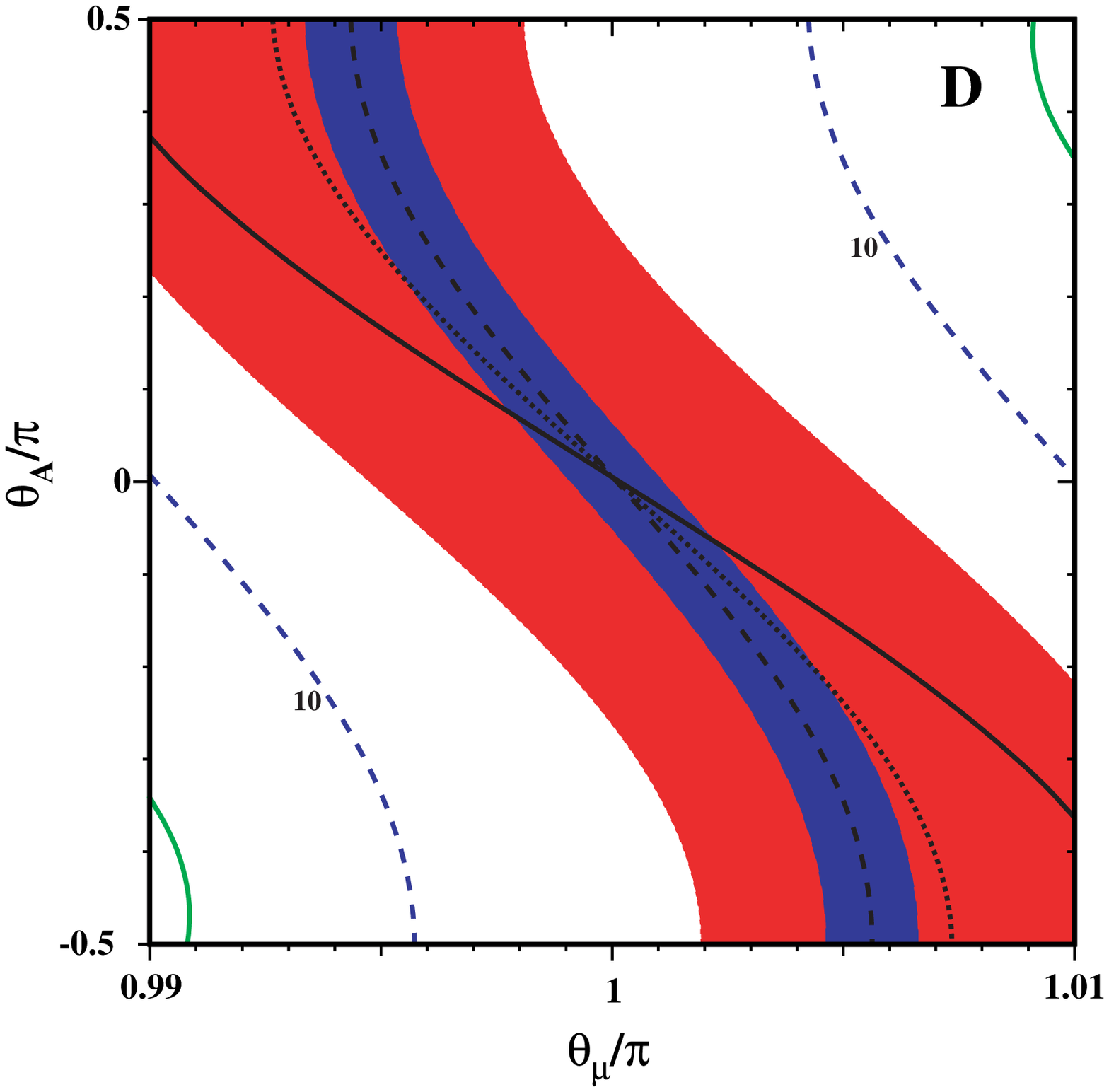}}
\centerline{\includegraphics[width=7cm]{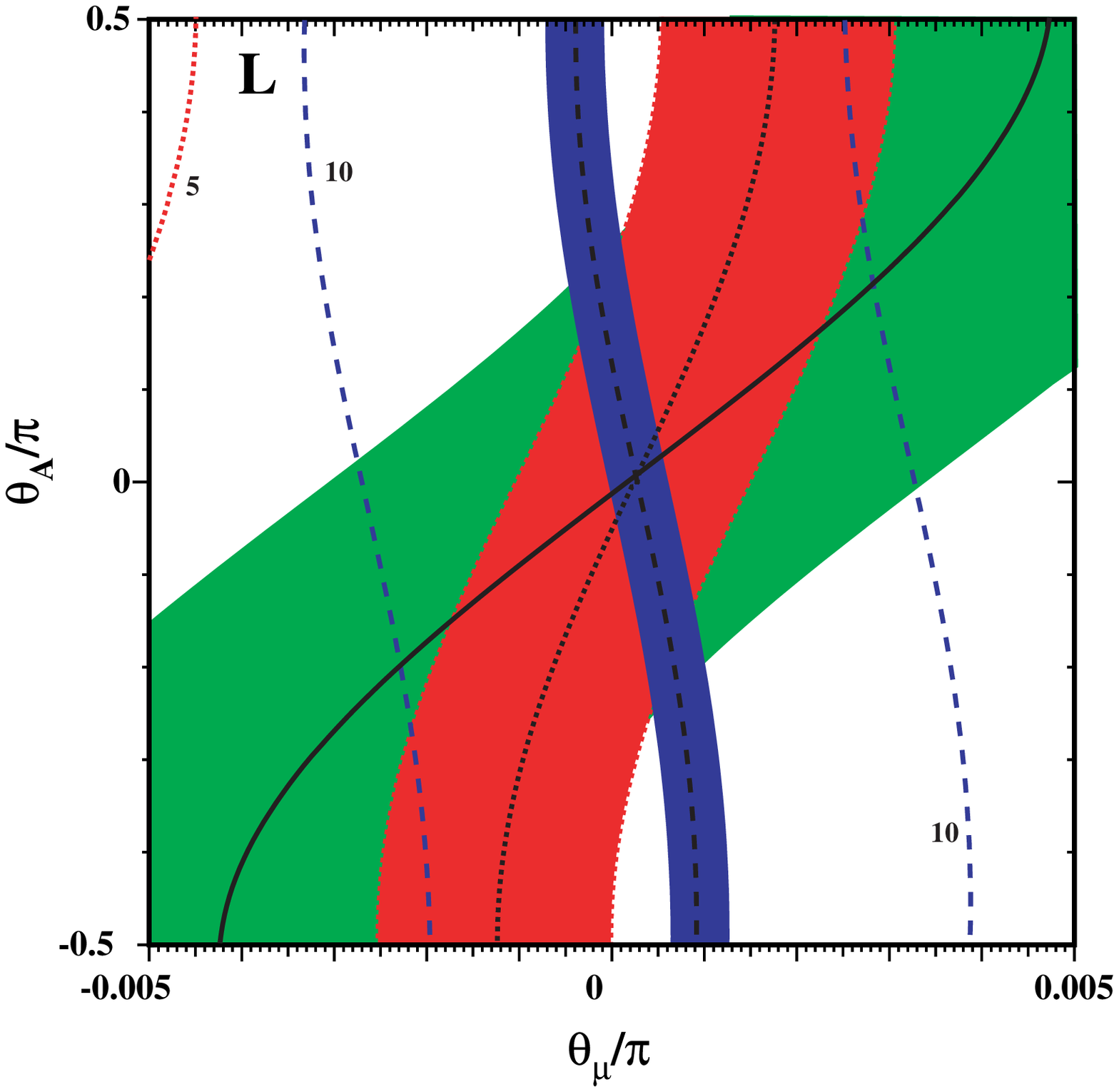}
\includegraphics[width=7cm]{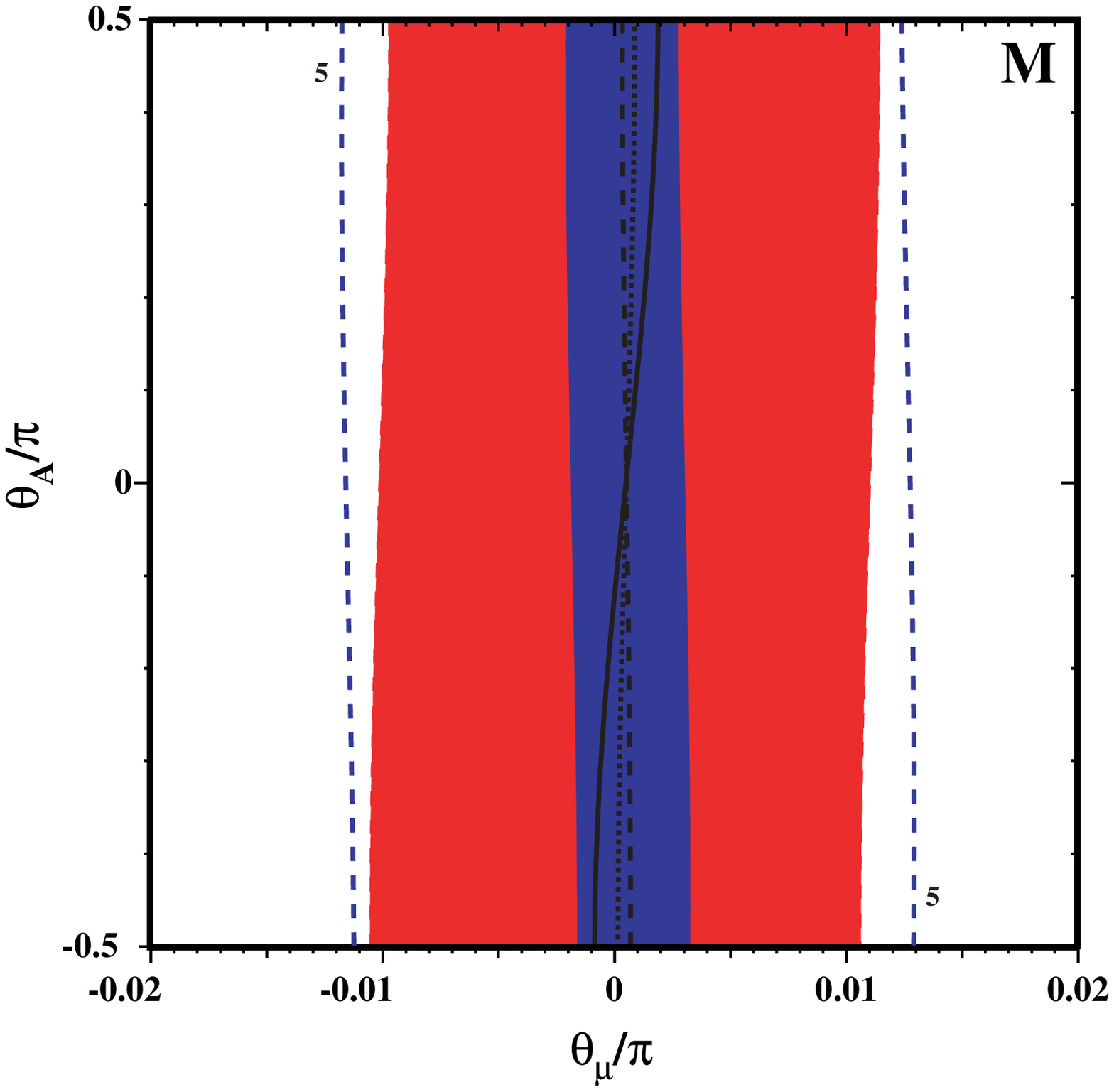}}
 \caption{\footnotesize The Tl (blue dashed), neutron (red dotted), and Hg
 (green solid) EDMs relative to their respective experimental limits in the 
 $\theta_\mu, \theta_A$ plane for benchmark points B, D, L and M.  
 Inside the shaded regions, the EDMs are less than
 or equal to their experimental bounds. Each of the EDMs vanish along the black contour
 within the shaded region.}
\label{bench} 
\end{figure}

We begin our analysis, therefore, with a discussion of EDMs in the context of 
four selected benchmark scenarios.  Point B is a bulk point with
$(m_{1/2}, m_0, \tan \beta) = (250, 75, 10)$.  The value of $m_0$
is shifted slightly ($+15$ GeV) from \cite{newbench} due the larger value of the 
top quark mass (we adopt $m_t = 178$ GeV as opposed to 175 GeV used previously)
and to several improvements
in the spectrum code, most notably the inclusion of the full set of 2-loop RGEs.
Note that here we have also chosen $A_0$ = 300 GeV as opposed to 0,
to allow for effects with non-zero phases. This also adds to the shift in $m_0$.
In Fig. \ref{bench}a, we show the magnitude of the EDMs for benchmark point B
in a plane defined by the phases of $A_0$ and $\mu$.
Within the blue shaded region the Tl EDM is less than or equal to its current 
experimental limit.  The blue (dashed) contours show the values of phases
when the Tl EDM exceeds its experimental bound by a factor of 10 and 50.
The black (dashed) curve within the shaded region has $d_{\rm Tl} = 0$. 
Similarly, within the red shaded region the neutron EDM is within its experimental bound.
The red dotted curve corresponds to a neutron EDM which is 10 times the experimental limit.
Finally, the green shaded region displays the Hg EDM bound and the green (solid) curves
shows where the Hg EDM is 5 times its experimental limit.  As expected, the bounds on 
$\theta_\mu$ are far stronger than those on $\theta_A$.  However,
because the EDMs cut the phase planes differently, when limits to all three EDMs 
are applied, a relatively strong limit ($|\theta_A/\pi| \la 0.08$) can be obtained in this case.
The limit on $|\theta_\mu|$ is approximately $0.002 \pi$.

In Fig \ref{bench}b, we show the result for benchmark point D, which is a
point along the $\chi-{\tilde \tau}$ co-annihilation tail and (in the absence of phases)
has $\mu < 0$ with $(m_{1/2}, m_0, \tan \beta) = (525, 130, 10)$, applying a 20 GeV shift in $m_0$.  
In this case, the sparticle masses are generally larger than found for point B, primarily
due to the increase in $m_{1/2}$. Here, we do not show the shaded region
obtained from the Hg EDM as it would fill the displayed plane. The Hg contours in the upper right
and lower left of the panel show where the EDM is equal to its experimental limit. For
point D, we find no limit to $\theta_A$, while $|\theta_\mu - \pi| \la 0.065 \pi$.

Point L (shown in Fig. \ref{bench}c) is also a  co-annihilation point, 
at large $\tan \beta$.  Point L is defined by
$(m_{1/2}, m_0, \tan \beta) = (450, 355, 50)$. In this case, because $\tan \beta$ is large,
we applied a larger shift (55 GeV) in $m_0$ to obtain the correct relic density when
phases are absent.  Once again, we recover a
bound on $\theta_A$, albeit a weak one of about $0.2 \pi$. In this case,
we have a very strong constraint on $\theta_\mu$, $|\theta_\mu| \la 0.0005 \pi$.
Finally, point M, shown in Fig. \ref{bench}d, is also a large $\tan \beta$ point found high in the 
funnel region with $(m_{1/2}, m_0, \tan \beta) = (1500, 1100, 57)$. Because the funnel
region is very  sensitive to $\tan \beta$, $m_{1/2}$ and $m_t$, 
we increased $\tan \beta$ from 50 to 57,
and lowered $m_{1/2}$ from 1840 GeV to 1500 GeV.  As for point D, 
we do not show the shaded region for the Hg EDM (as it would fill the plane as displayed).
As one can see, there is no bound on $\theta_A$, and despite the large values of $m_{1/2}$ and
$m_0$, there is still a significant bound on $|\theta_\mu|$ of 0.004$\pi$. 

The constraints for all four benchmark scenarios are summarized in Table~1.

\begin{table}
\centerline{
\begin{tabular}{||c|c|c||} \hline\hline
 benchmark point ($m_{1/2},m_0,\tan\beta$) & $|\th_A/\pi|$ & $|\th_\mu/\pi|$ \\ \hline
 B (250,75,10) & $\la 0.08$ & $\la 0.002$ \\
 D (525,130,10) & - & $\la 0.07$ \\
 L (450,355,50) & $\la 0.2$ & $\la 0.0005$ \\
 M (1500,1100,57) & - & $\la 0.004$ \\ \hline\hline
\end{tabular}}
\caption{\footnotesize A summary of the constraints on the phases within the four benchmark scenarios
shown in Fig.~\ref{bench}. All points use $|A_0|=300$ GeV, and the quoted bound on the
phase of $\mu$ for point $D$ is modulo $\pi$.}
\end{table}

\subsubsection{General constraints on the CMSSM parameter space}

We now turn to a more general analysis of the CMSSM ($m_{1/2}, m_0$) plane
for fixed values of $\tan \beta$. We will present the results in several different ways.
One motivation for restricting our attention to the CMSSM is that we can subject the theory
to a full suite of different phenomenological constraints. On the following parameter space plots
we will indicate the excluded and preferred regions due to the following:

\begin{itemize}
\item {\bf LSP dark matter:} Regions in parameter space where the LSP is charged, generically
a stau, are not viable and can be excluded. Moreover, if one demands that the relic density is consistent
with the range $\Omega h^2 = 0.0945 - 0.1287$ determined by WMAP \cite{wmap}, this limits 
the viable region to
thin strips in the $(m_{1/2},m_0)$ plane. Furthermore, the relic density is generically 
too large on one side of the WMAP strip. We will calculate this relic density in the absence of
phases, but will restrict the magnitude of the phases so that their impact
should be negligible \cite{WWII,fo2}.
\item {\bf Contributions to $b\rightarrow  s\ga$:} Within the flavor sector, constraints on 
the supersymmetric contributions to $b\rightarrow  s\ga$ are often important for low values
of $m_{1/2}$, particularly at large $\tan \beta$ \cite{bsg}. 
\item {\bf LEP chargino and Higgs searches:} The constraints resulting from the null LEP results
for Higgs \cite{Higgs} and chargino \cite{Chargino} searches also impinge on the CMSSM parameter 
space for small $m_{1/2}$.
\end{itemize}

Our motivation is not to search specifically for regions allowed by all of these constraints, but to 
emphasize the complementarity of the constraints, and the relative importance of those imposed by EDMs.
The resulting conclusions should then be relevant, qualitatively at least, in more general SUSY-breaking
scenarios.

In presenting our results, we will consider the contribution of the two phases separately, and
we will also use two phase values of $\pi/20$ and $\pi/6$, and show contour plots of $d/d_{\rm exp}$,
the EDM relative to its current experimental bound (or future level of sensitivity). These phases
are both within the linear regime and thus the contours can easily be converted to bounds on 
the phases, $\th \la \th_0/ (d/d_{\rm exp})$, where $\th_0$ denotes the nonzero phase chosen
for the plot. This allows for a simple rescaling of our results given updated future limits.

The smaller phase value, $\pi/20$, was chosen to restrict attention the phenomenologically
most interesting region, where the spectrum is not highly tuned. Note that the tuning of this
phase is comparable to the minimal tuning that must be imposed on the spectrum in the CMSSM, 
as implied by the LEP bounds and the electroweak precision tests (see e.g. \cite{ewpts} for a recent 
analysis within the MSSM). We have also included results
with a ``near maximal'' phase, $\pi/6$, to illustrate the full reach of the EDM constraints
with regard to the parameter space of the CMSSM and the SUSY spectrum. The value $\pi/6$ was
adopted as the boundary of the linear regime, to allow for a simple interpretation of the contours,
and, if viewed as a ``maximal'' phase, also takes into account a conservative (factor of two) 
uncertainty in some of the hadronic EDMs.

We will now present the results in more detail.

\bigskip
{\it Low $\tan\beta$ and future sensitivity}
\bigskip

\begin{figure}
\centerline{\includegraphics[width=7cm]{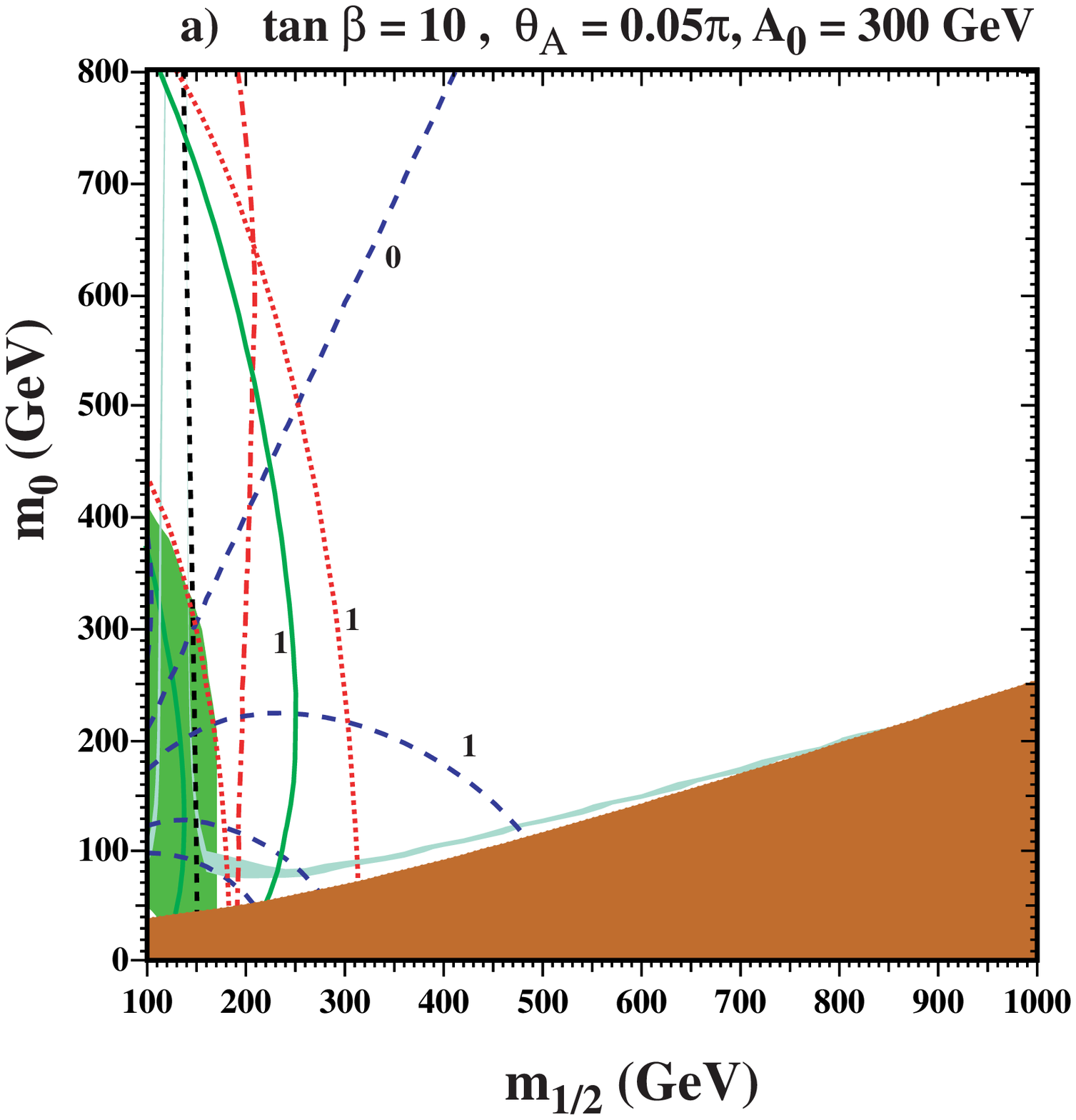}
\includegraphics[width=7cm]{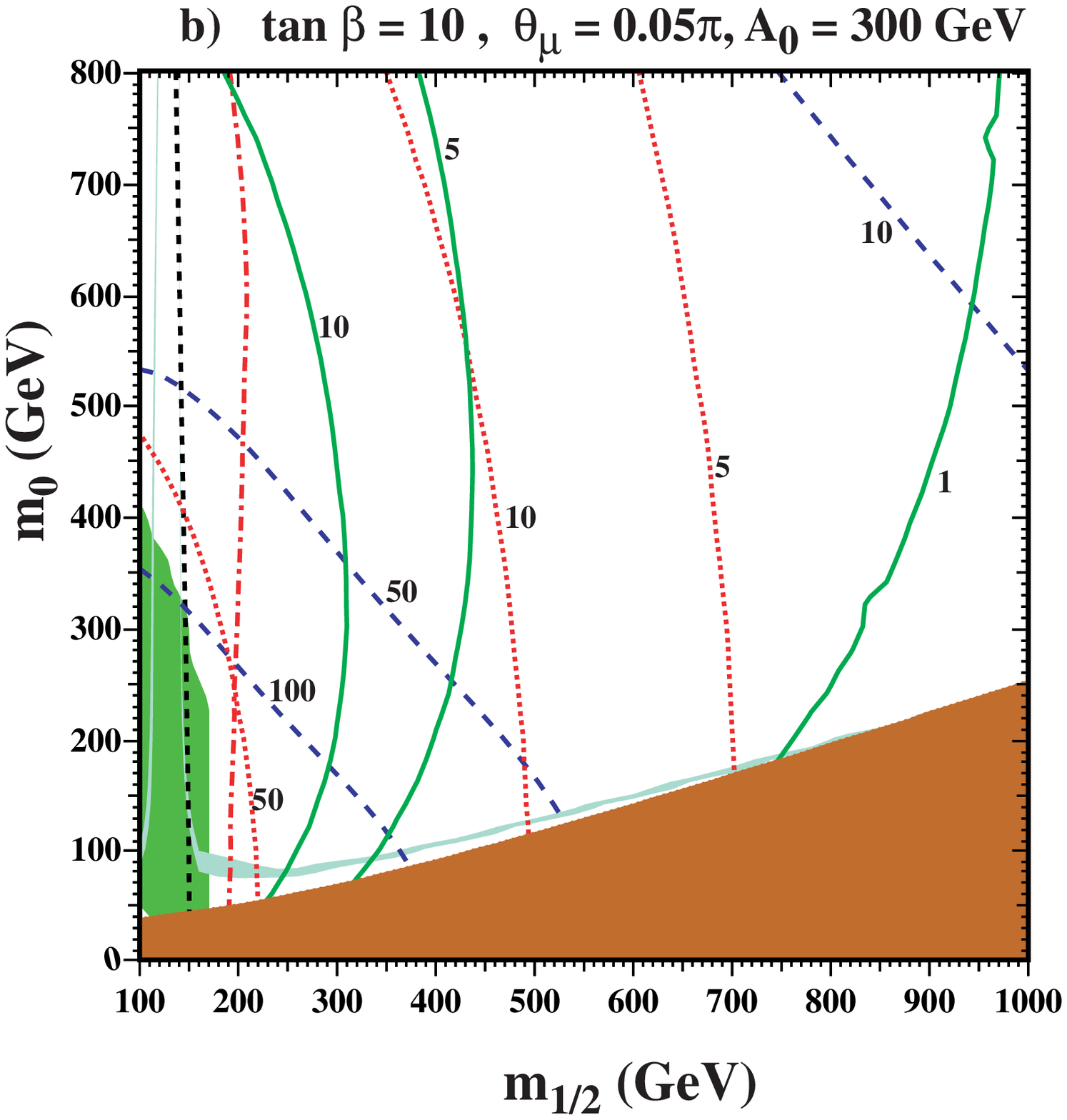}}
\centerline{\includegraphics[width=7cm]{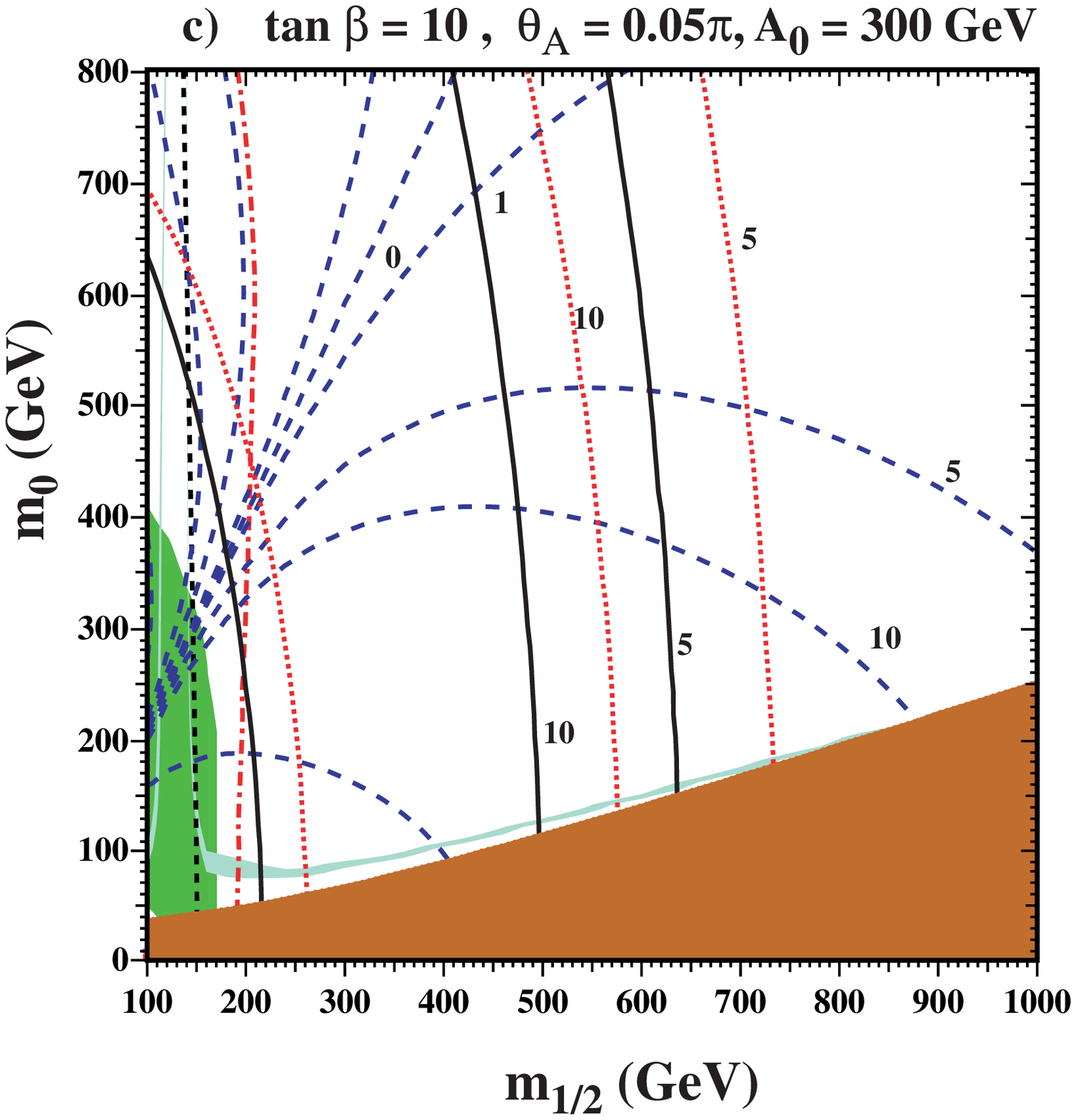}
\includegraphics[width=7cm]{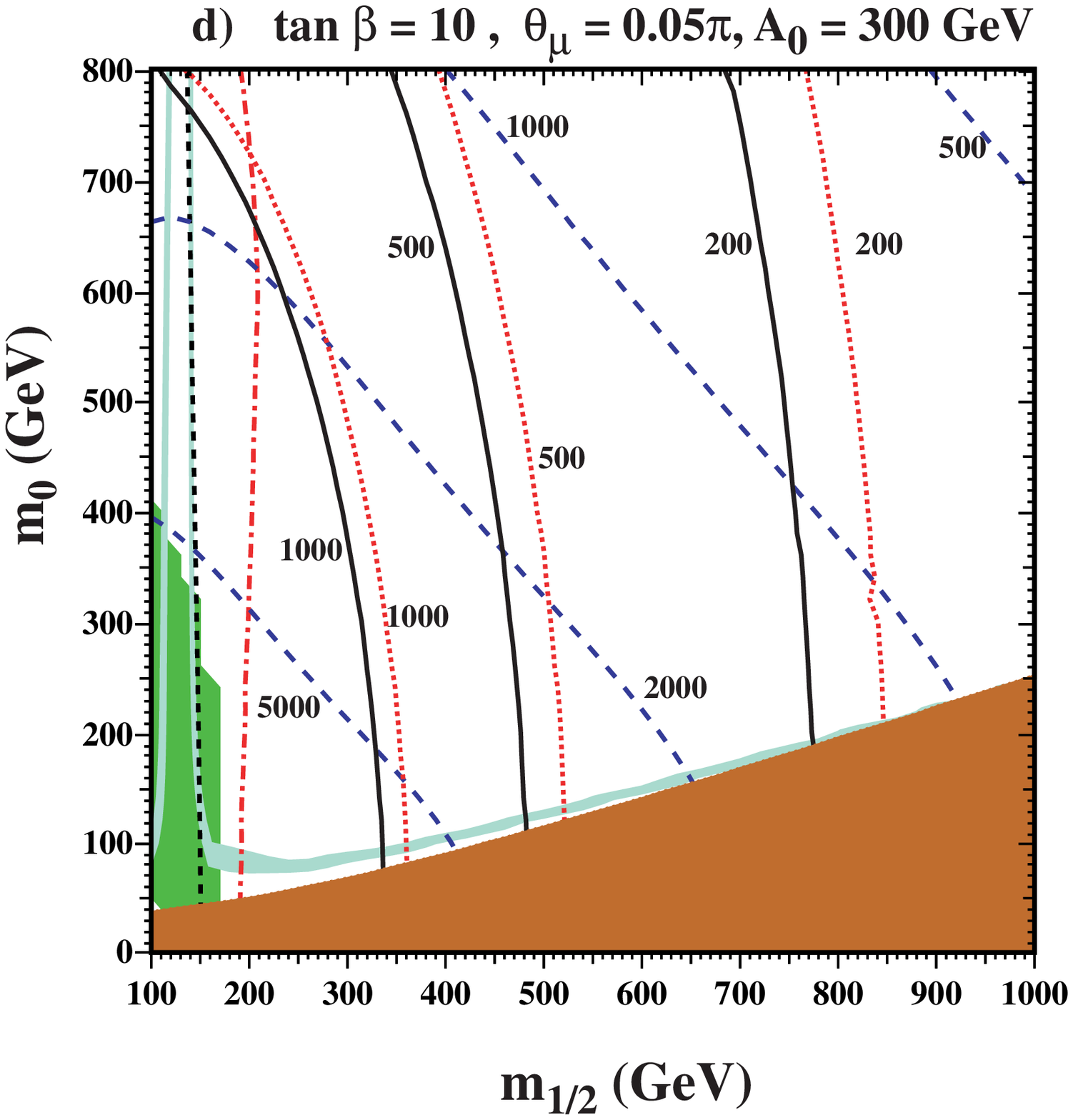}}
 \caption{\footnotesize The Tl (blue dashed), neutron (red dotted), and Hg
 (green solid) EDMs relative to their current experimental limits (a,b) and 
 the electron (blue dashed), the neutron (red dotted), and the deuteron (black solid)
 EDMs relative to possible future experimental limits (c,d) in the 
 $m_{1/2}, m_0$ plane of the CMSSM. 
 In each panel, $|A_0| = 300$ GeV with $\tan \beta = 10$.
 The shaded regions are described in the text. In a) and c), $\theta_A = 0.05\pi$,
 and in b) and d), $\theta_\mu = 0.05\pi$. }
\label{tb10} 
\end{figure}

In Fig. \ref{tb10}a,b, we show the current constraints in the CMSSM
plane for $\tan \beta = 10$ due to the Tl, neutron, and Hg EDMs for $\theta_A = 0.05\pi$ (a)
and $\theta_\mu = 0.05\pi$ (b).  As in Fig. \ref{bench}, the blue (dashed), red (dotted), and green
(solid) curves correspond to the Tl, neutron, and Hg EDMs respectively.  As described above, the contour
labels show the magnitude of the EDM relative to the current experimental bound.
Throughout the $(m_{1/2},m_0)$ plane, we have taken $m_t = 178$ GeV, $m_b(m_b) = 4.25$ GeV, and 
$|A_0| = 300$ GeV, with $\theta_A = 0.05\pi$, $\theta_\mu = 0$ in a), and $\theta_\mu = 0.05\pi$, 
$\theta_A = 0$ in b).  In each of the panels, the brown shaded  region in the lower right 
corresponds to that portion of the CMSSM plane where the right-handed stau is the LSP,
and is therefore excluded.  The green shaded region in the left of each panel is
also excluded as there the supersymmetric contribution to $b \to s \gamma$ is
too large. Also shown are the current constraints from LEP based on the masses
of the chargino (nearly vertical black dashed line) and Higgs boson (red dot-dashed).
Regions to the left of these curves are excluded.  The grey shaded wisp-like region,
corresponds to that portion of the plane where the cosmological relic density lies in the range 
$\Omega h^2 = 0.0945 - 0.1287$ as determined by WMAP. Any area above this shaded region is excluded 
as the relic density is too large. Note that the very thin area {\em below} the WMAP strip and above the
stau LSP region is allowed  so long as there is another source for dark matter.

In Fig.~\ref{tb10}a, we show the contours for which each of the 
three EDMs considered, saturates its experimental bound (curves labeled 1).
Regions to the left of (or below) these curves are excluded.  The constraints from $d_n$ and 
$d_{\rm Hg}$ are relatively uniform. However, we see that for the Tl EDM, there is a line through the 
plane where a cancellation occurs, and $d_{\rm Tl} = 0$. Interestingly enough, despite the
relatively low value for $\tan\beta$, this cancellation line can be understood through the
impact of 2-loop RG corrections to Im$M_i$. 

To see this, recall that for low $\tan\beta$ the Tl EDM is dominated by 1-loop threshold corrections
to $d_e$, which to a reasonable approximation take the form shown in Eq.~(\ref{simplified}). Since
$\mu$ is real, the {\it negative} RG corrections to Im$M_2$ can lead to a cancellation between
the two leading contributions. From (\ref{estimate}),
\be
d_e  \propto \left[ 5g_2^2\tan\beta \mu {\rm Im} (M_2) 
+ 2g_1^2 {\rm Im} (M_1^* A_e)\right]\!,
\label{cancel1}
\ee
using the leading-log estimate for Im$M_2$ in (\ref{estimate}), and ignoring 
the subleading imaginary correction to $M_1$, we can estimate the condition for dominance
of the first term in (\ref{cancel1}) as
\be
 \tan\beta \frac{\mu {\rm Im}(A_0)}{M_1 {\rm Im}(A_e)} \ge 17 \;\;\;\; \Longrightarrow \;\;\;\;
  \tan\beta \ge 7,
 \label{cancel}
\ee 
where in the second relation we have used 
$|\mu|\sim (-m_{H_2}^2)^{1/2} \sim 1.6\, m_{1/2}$, $M_1\sim 0.4\, m_{1/2}$,
and Im$(A_e)\sim 1.6\,$Im$(A_0)$ to provide an indication of the size of $\tan\beta$ needed to
achieve a cancellation in the relevant part of the $(m_{1/2},m_0)$ plane. We see 
that moderate values of $\tan\beta$ are already sufficient,
and indeed this cancellation is apparent in Fig.\ref{tb10}a. Its precise shape depends on a number
of other contributions that we have neglected, which become more important at large $\tan\beta$ 
as we will discuss below.  For 
completeness, the plot also indicates contours for the EDMs which are 5 times their 
experimental bound, and one contour for Tl where the EDM is 10 times it experimental value 
(the innermost of the blue dashed curves).  

In contrast, when $\theta_\mu = 0.05\pi$, as in Fig. \ref{tb10}b, 
the constraints become much stronger.  All the curves are now labelled by the value
of $d/d_{\rm exp}$, i.e. the ratio by which the predicted EDM exceeds the experimental bound.
For this choice of $\theta_\mu$, we see that the entire region of the plane shown in the Figure is 
excluded. The Tl EDM, which now exhibits no cancellation line as it is dominated by the first term
in (\ref{simplified}), supplies by far the strongest constraint. Nonetheless, $d_n$ and $d_{\rm Hg}$ 
also provide strong constraints across the plane. As described above, since the phase is small
and the contribution to the EDM is thus linear, we can also use the curves to place a limit 
on $\theta_\mu$.  For example, the region to the right of  the Tl curve labeled 50 is allowed so 
long as $\theta_\mu \le 0.05\pi/50 = 0.001\pi$. Similarly for the other curves in this panel.

In the lower half of Fig.~\ref{tb10}, we consider how the present situation will be modified 
with the next generation of EDM experiments, and exhibit the anticipated sensitivities.
In Fig. \ref{tb10}c and d, we show $d_e$ (blue dashed), the neutron (red dotted),
and the deuteron (black solid) relative to the anticipated future level of sensitivity  as 
described in section 3.4. We emphasize that there are several experiments in this list sensitive
to $d_e$, and we have simply deduced the sensitivity to $d_e$ from these, ignoring corrections
from electron-nucleon interactions. For small values of $\tan\beta$ as we have here, this should
not be a significant problem but one should bear in mind that these conclusions cannot easily be
extrapolated to other regimes, e.g. to large $\tan\beta$.

In panel c), the contours shown are 5, 10, and 50,
where the latter are unlabeled. For $d_e$, contours of 0 and 1 are also shown.
As one can see we expect the deuteron and neutron to give comparable limits
which are quite complementary to that obtained from the future bound on the electron 
EDM from PbO, YbF, etc. As is apparent from the comparison of Fig.~\ref{tb10}a and  Fig.~\ref{tb10}c,
the limits on the SUSY parameter space are expected to be dramatically improved by future 
experiments, particularly so for the dependence on $\th_A$.  For $\theta_A = 0.05 \pi$, 
the entire (displayed) plane would be excluded by both the n and D EDMs, provided 
that future experiments will see zero EDMs, thus forcing a 
smaller value for the phase of $A_0$.  In panel d), we see that the $d_e$ limits become so strong 
that even at the endpoint of the co-annihilation region at $m_{1/2} \simeq 900$ GeV, 
we are forced to very small phases, $\theta_\mu \la 5\times 10^{-5}\pi$.

\bigskip
{\it Constraints for large $\tan\beta$}
\bigskip

\begin{figure}
\centerline{\includegraphics[width=6cm]{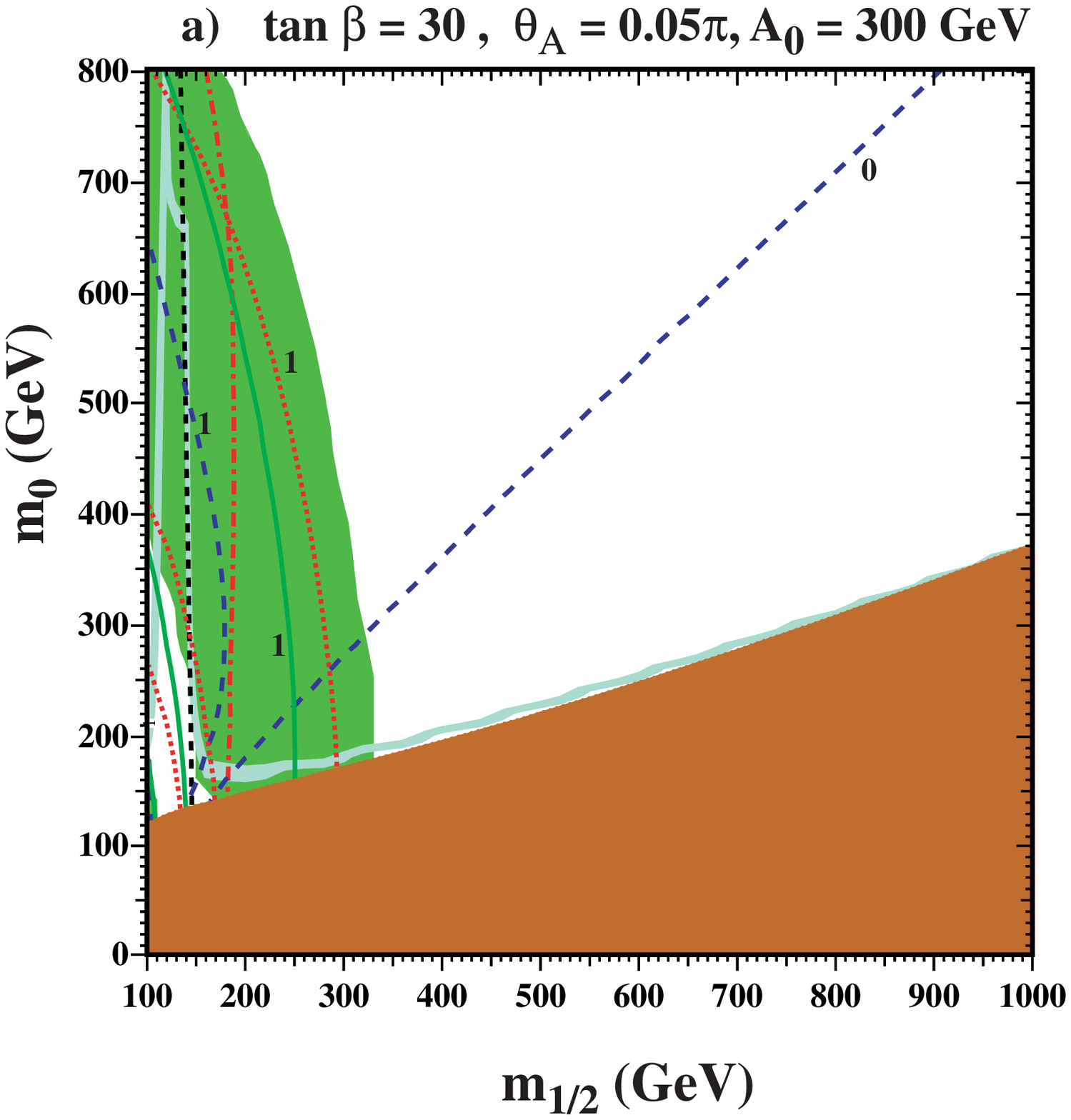}
\includegraphics[width=6cm]{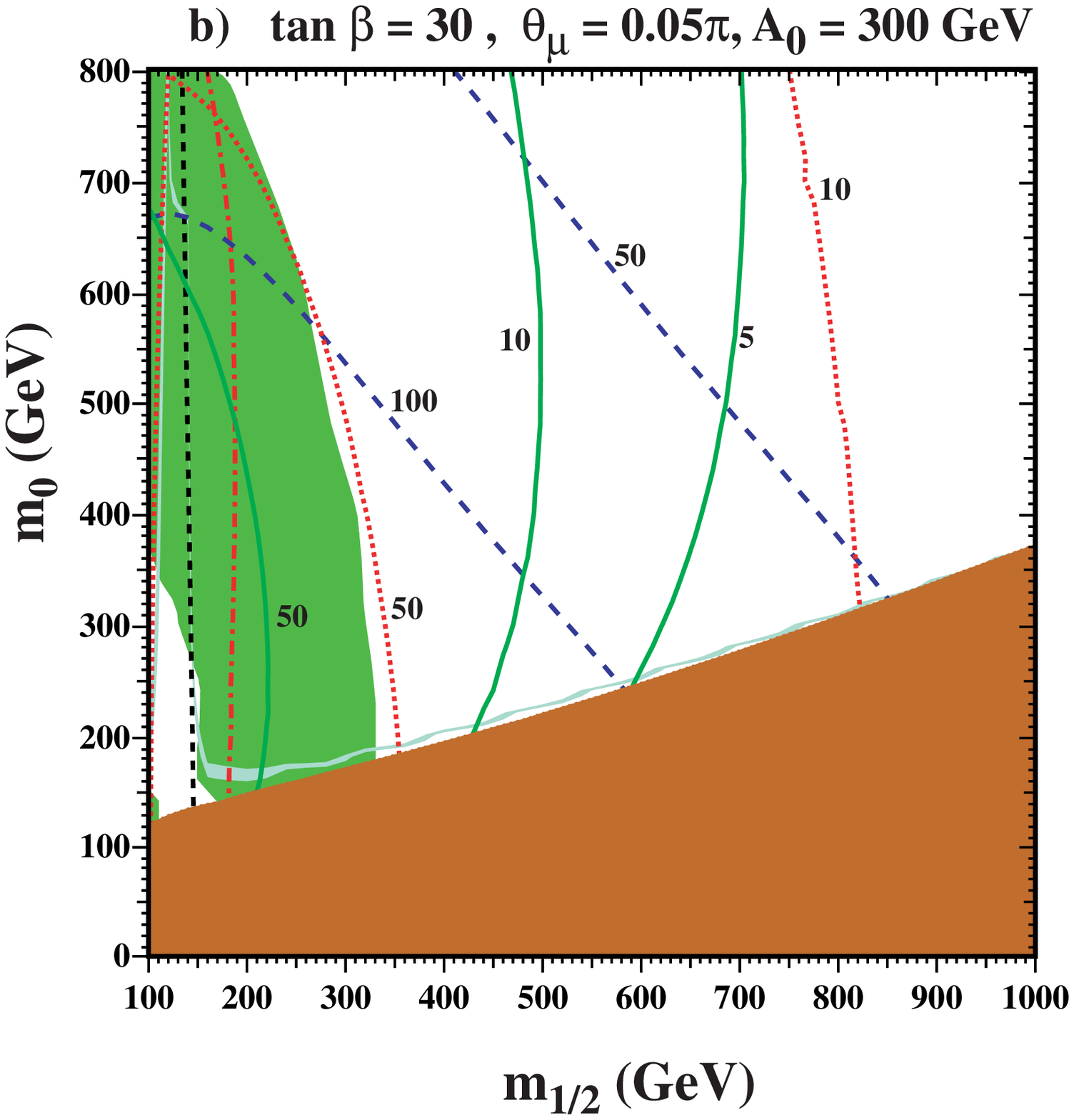}}
\centerline{\includegraphics[width=6cm]{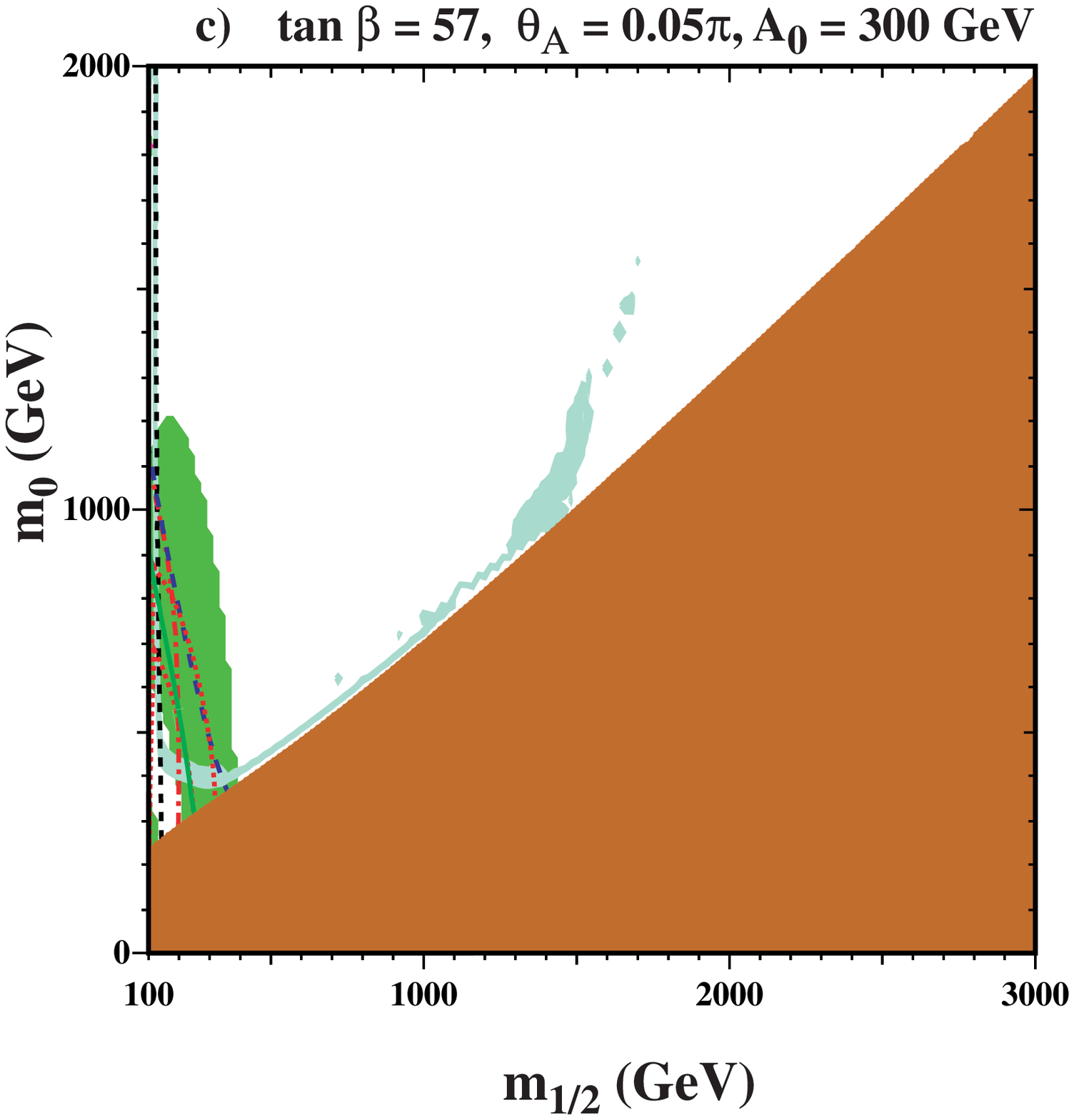}
\includegraphics[width=6cm]{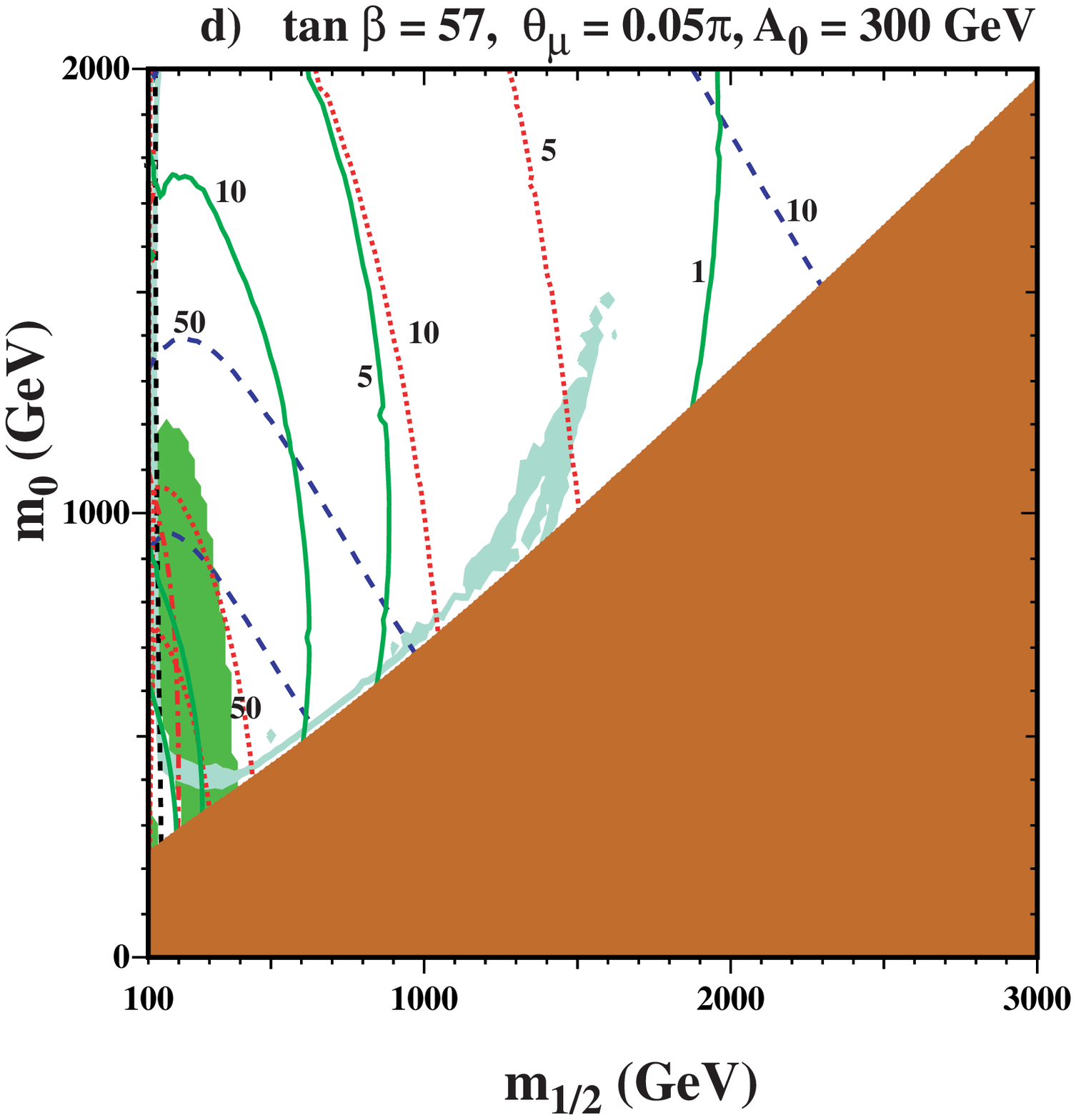}}
\centerline{\includegraphics[width=6cm]{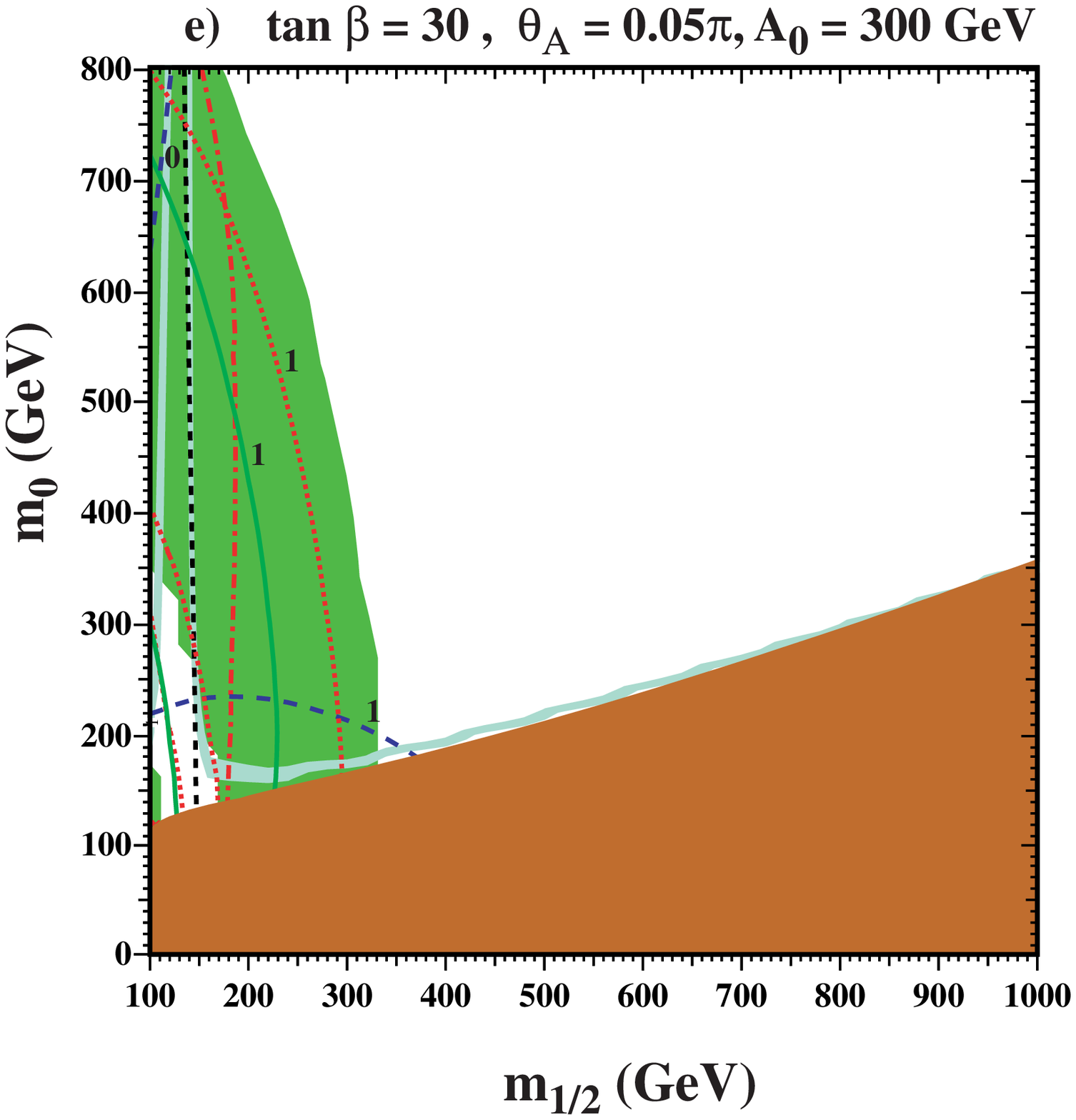}
\includegraphics[width=6cm]{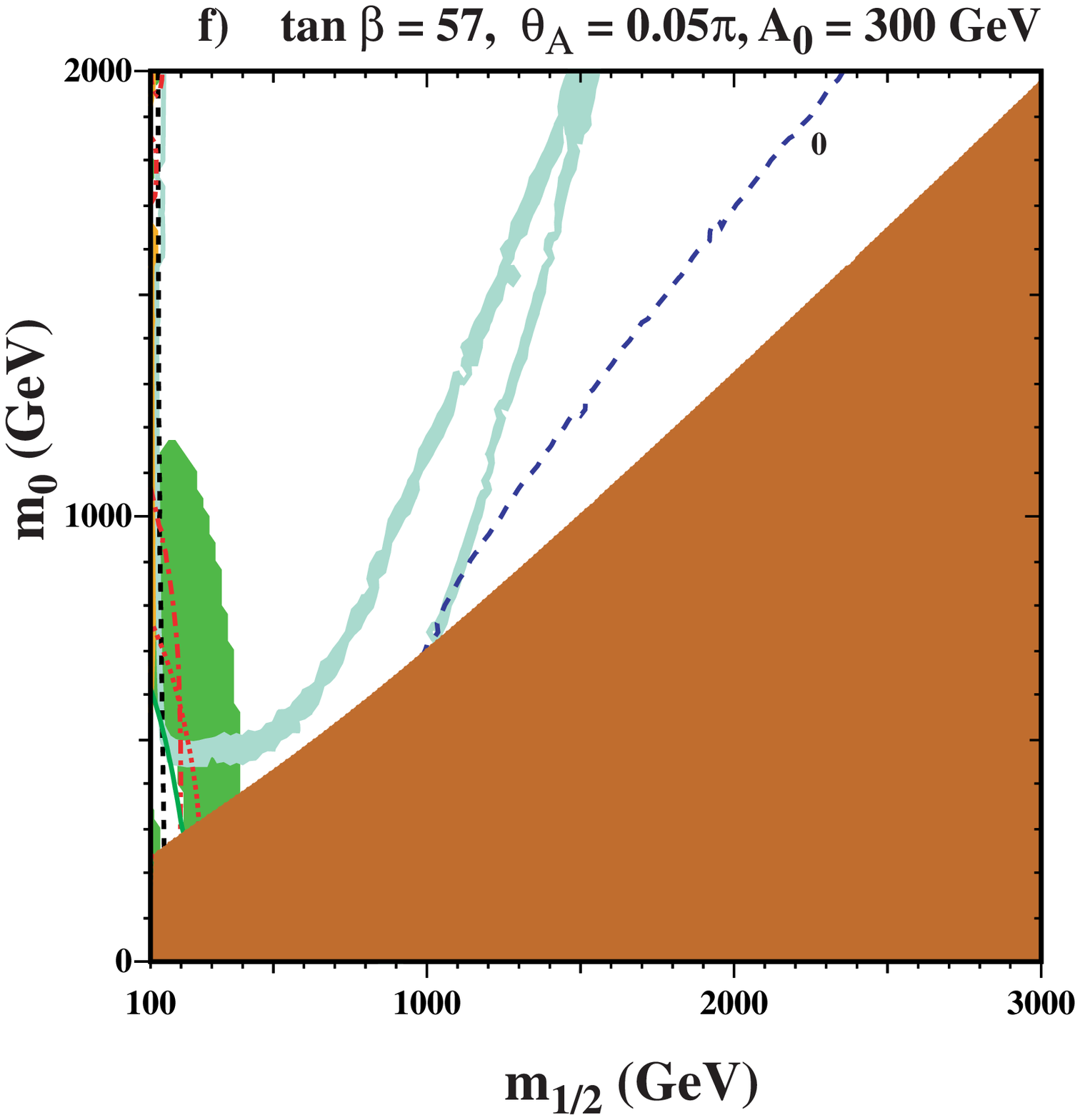}}
 \caption{\footnotesize As in Fig. \protect\ref{tb10}, the Tl (blue dashed), neutron (red dotted), and Hg
 (green solid) EDMs relative to their current 
 respective experimental limits in the 
 $m_{1/2}, m_0$ plane of the CMSSM for $\tan \beta = 30$ (a, b, and e), and 
 $\tan \beta = 57$ (c, d, and f). 
 In each panel, $|A_0| = 300$ GeV.
 In a, c, e, and f), $\theta_A = 0.05\pi$,
 and in b) and d), $\theta_\mu = 0.05\pi$. In panels e and f) only 1-loop RGEs were used to 
 compute the supersymmetric particle spectrum.}
\label{hightb} 
\end{figure}

The large $\tan\beta$ regime exhibits a number of new qualitative features as concerns the
dependence on the underlying sources. The 1-loop $\th_\mu$--dependence of the down-type fermions
is linearly enhanced, while as we have discussed RG corrections to Im$M_i$ can similarly
provide significant corrections for the dependence on $\th_A$. A number of 2-loop Barr-Zee type
contributions \cite{ckp,pilaftsis02} to the EDMs and CEDMs are also $\tan\beta$-enhanced and 
become significant
in this regime, while the largest qualitative change arises through the fact that various four-fermion
operators provide sizable contributions \cite{Barr,LP,dlopr}, 
and compete with the constituent EDMs. The most prominent
example of this is the contribution of $C_S$ to $d_{\rm Tl}$ as described in Section~3. A comprehensive
study of the relative importance of these contributions was carried out in \cite{dlopr}, and here
we will present combined results for the EDMs within the framework of the CMSSM. It is apparent
from the results of \cite{dlopr} that the restrictive ansatz of the CMSSM does not exhibit all 
the possible regimes that may arise in a more general MSSM scenario, and new qualitative features
are restricted primarily to the behaviour of $d_{\rm Tl}$. 

Our results are presented in the upper two rows of Fig.~\ref{hightb}, for
$\tan\beta=30$ and $57$ respectively, and again for cases with either nonzero $\th_A=0.05\pi$
or $\th_\mu=0.05\pi$. As mentioned, the hadronic EDMs, $d_n$ and $d_{\rm Hg}$, are still primarily
dominated by the one-loop contributions and thus are not altered significantly for 
nonzero $\th_A$ (cf. Fig.~\ref{hightb}a,c with Fig.~\ref{tb10}a), while they scale up 
linearly with $\tan\beta$ for nonzero $\th_\mu$ (cf. Fig.~\ref{hightb}b,d with Fig.~\ref{tb10}b).
We recall that although the 2-loop RG-induced corrections for nonzero $\th_A$ are present
and $\tan\beta$-enhanced, they are not particularly large for the quark EDMs and CEDMs as
all contributions arise predominantly via squark-gluino loops. Indeed, it is apparent
from Fig.~\ref{hightb}a and c that, for our rather small value of $\theta_A = 0.05 \pi$,
the EDMs are less constraining than the limits from $b \to s \gamma$. On the plots
for $\tan \beta = 57$, one may note the appearance of the funnel-like region where 
dark matter annihilations are mediated by $s$-channel heavy Higgs exchange.

Let us now focus attention on the contours for $d_{\rm Tl}$. At first sight, the results
for $\tan\beta=30$ may not look dramatically different from those for $\tan\beta=10$. We see once again
a cancellation line extending out now to somewhat larger values of $m_{1/2}$. However, its
precise form is now determined by a more complex interplay between many different effects.
Indeed, from (\ref{cancel}) we might
expect that the Im$M_2$-induced corrections would now dominate the 1-loop contribution to $d_e$. This
is indeed the case, but for $\tan\beta=30$ the 2-loop corrections and the contributions from $C_S$
are also competitive, and the observed cancellation line results from the interplay of all of these sources.
It is only for sufficiently large $\tan\beta$, as observed for $\tan\beta=57$ in Fig.~\ref{hightb}c,
that the cancellation line disappears, as the various contributions fall out of balance, although
there are still some partial intermediate cancellations \cite{dlopr}.

We have not included plots outlining the future sensitivity in this case, due to the lack of
atomic calculations detailing the contributions from four-fermion operators for the relevant
paramagnetic sources. Nonetheless, by comparison with Fig.~\ref{tb10}c it should be clear that
much of the displayed plane will again be covered by the EDM reach, even for $\th_A=0.05\pi$, 
providing a significant improvement over the current results in Fig.~\ref{hightb}a and c.

\bigskip
{\it Impact of two-loop RG-evolution of the phases}
\bigskip

To gauge the effect of the inclusion of the 2-loop RGEs, we have also shown in Fig.~\ref{hightb}e and f 
the constraints on the CMSSM plane when only the 1-loop RGEs are run for non-zero
$\theta_A$. For clarity, we have only displayed contours with $d = d_{\rm exp}$. As anticipated
in our discussion in Section~2, the Tl EDM is most affected by the inclusion of the 2-loop RGEs
and as one sees, the cancellation line at $\tan \beta = 30$, which using the 1-loop RGEs barely appears
on the top left corner of the plot, is rotated clockwise and shifted to close to the middle of the
plane. Similarly, for $\tan \beta = 57$, the cancellation that is apparent when using the 1-loop
RGEs is shifted into the stau LSP region in Fig.~\ref{hightb}c. 

One should bear in mind in viewing these plots that the spectrum is also slightly perturbed
in shifting from 1-loop to 2-loop RGEs, but this is a minor effect and the dominant change in
the EDMs is due to the induced corrections to Im$M_i$. The significant shift in the 
WMAP band is due to the fact that the funnel region is highly sensitive to the pseudoscalar
Higgs mass.

\bigskip
{\it Full EDM reach in the CMSSM parameter space}
\bigskip

In the analysis above, we have chosen rather small phases, consistent with the level of
implied tuning in the CMSSM, to focus in on the phenomenologically most interesting
region. However, it is also clearly of interest to determine precisely how far the EDMs can
reach in terms of the parameter space of the CMSSM, and more generally in terms of the 
sparticle spectra. Thus, we will now boost the phases to $\pi/6\sim 0.17\pi$, which we will
take as the boundary of the linear domain, so that we can still use the figures to approximate the limit on
the given phase at any specific point in the CMSSM plane. As discussed earlier, if $\pi/6$ is 
taken as a maximal phase value, it also provides a conservative treatment of some of the 
calculational uncertainties. 

\begin{figure}
\centerline{\includegraphics[width=7cm]{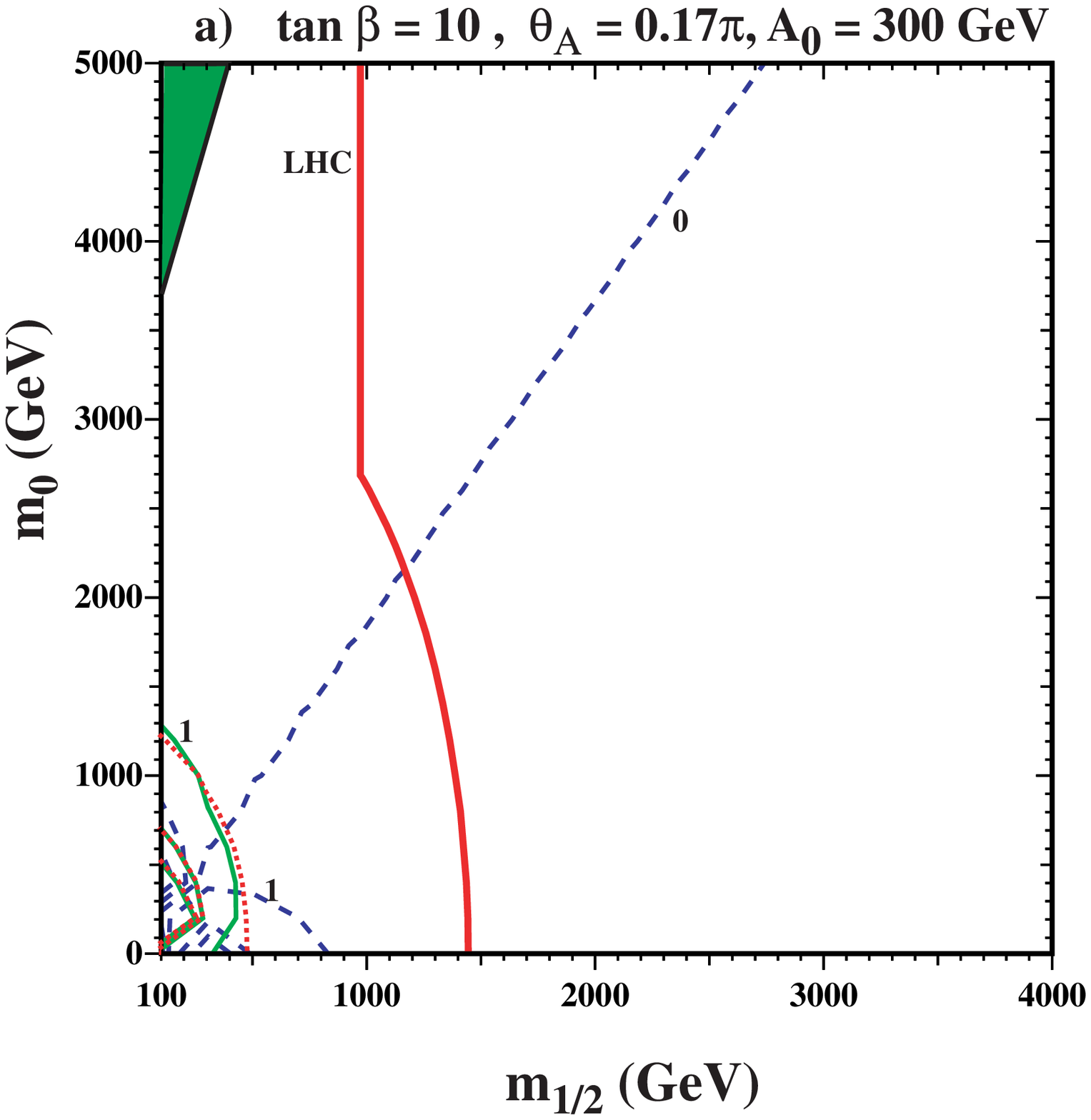}
\includegraphics[width=7cm]{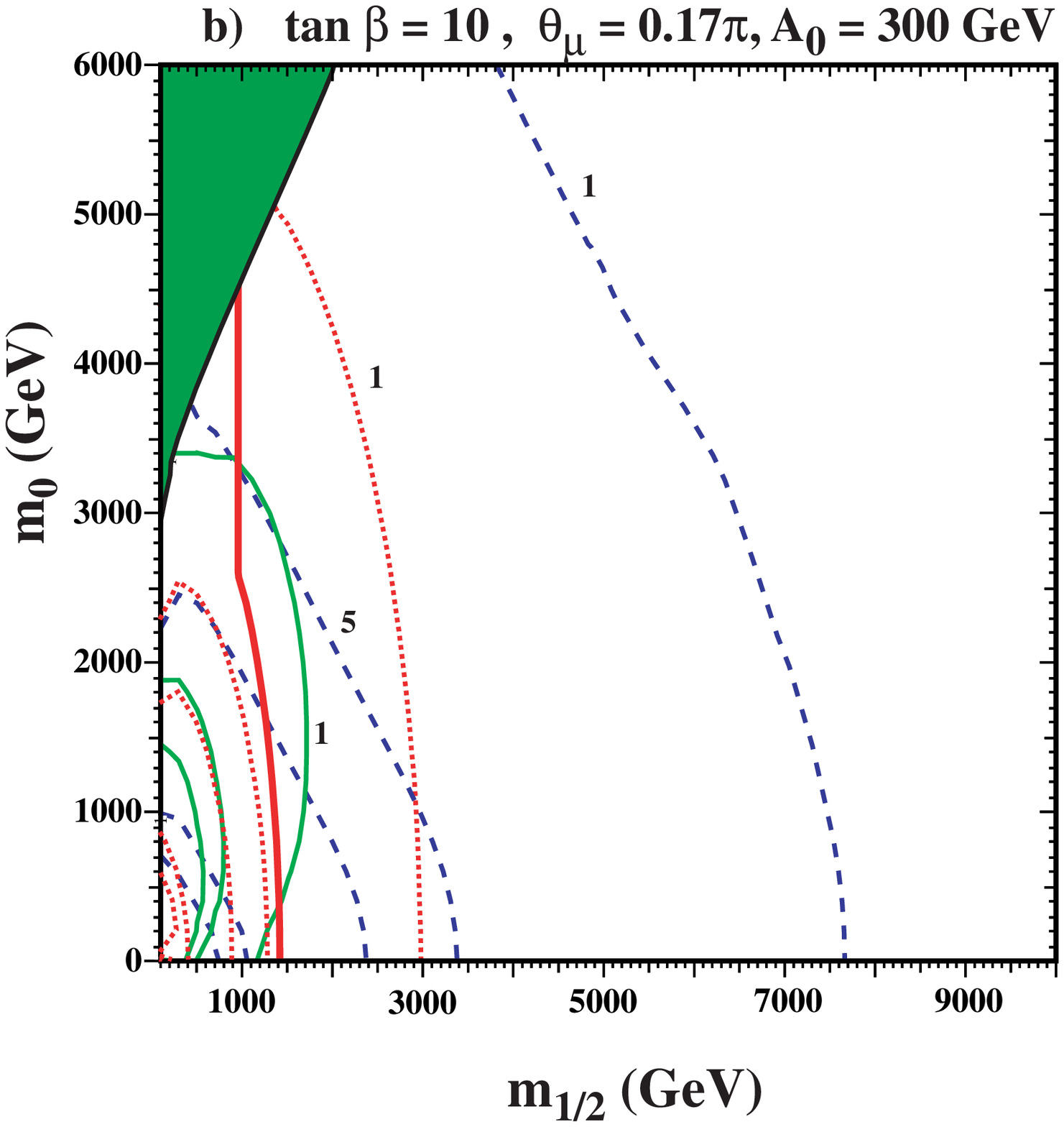}}
\centerline{\includegraphics[width=7cm]{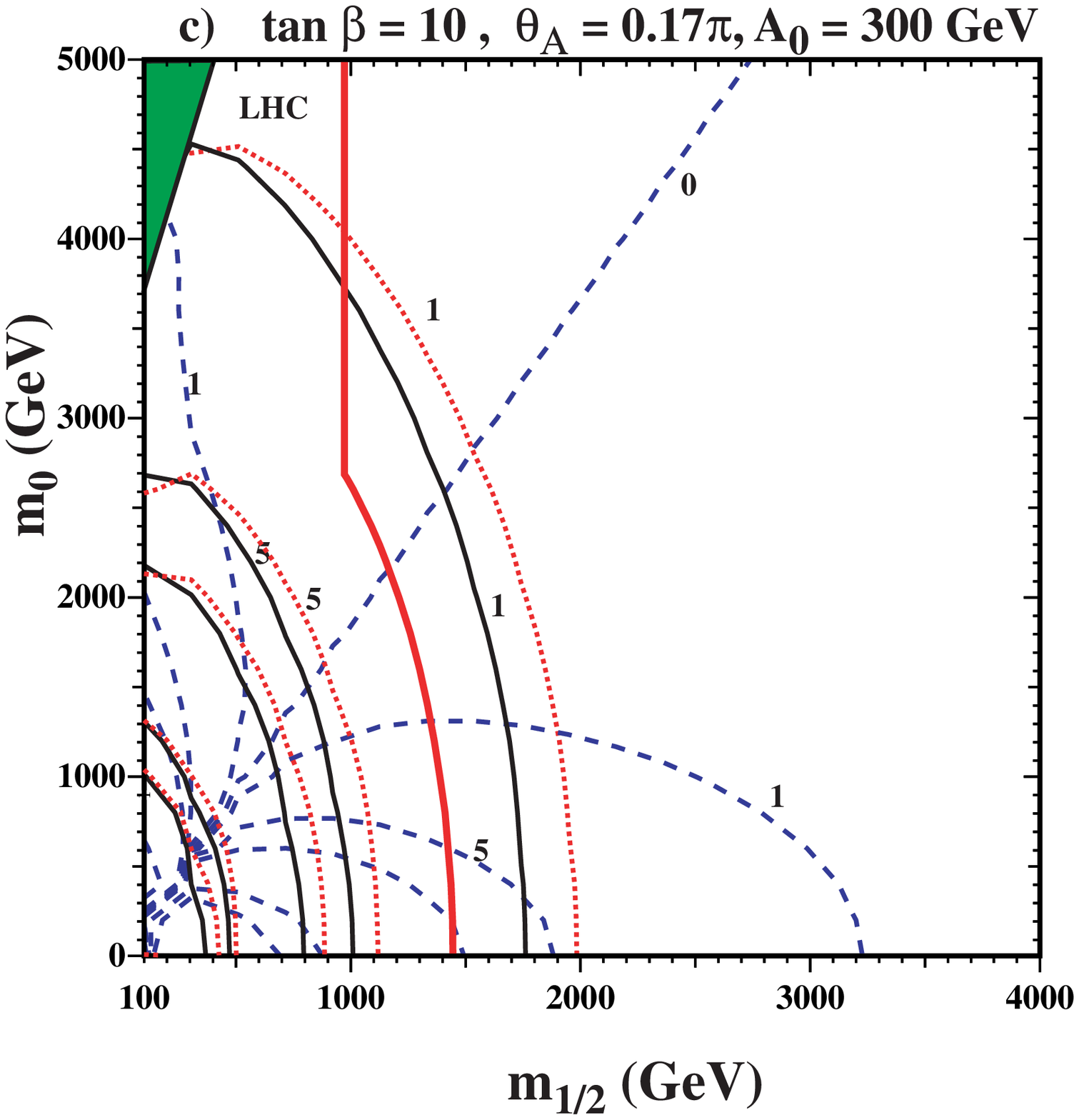}
\includegraphics[width=7cm]{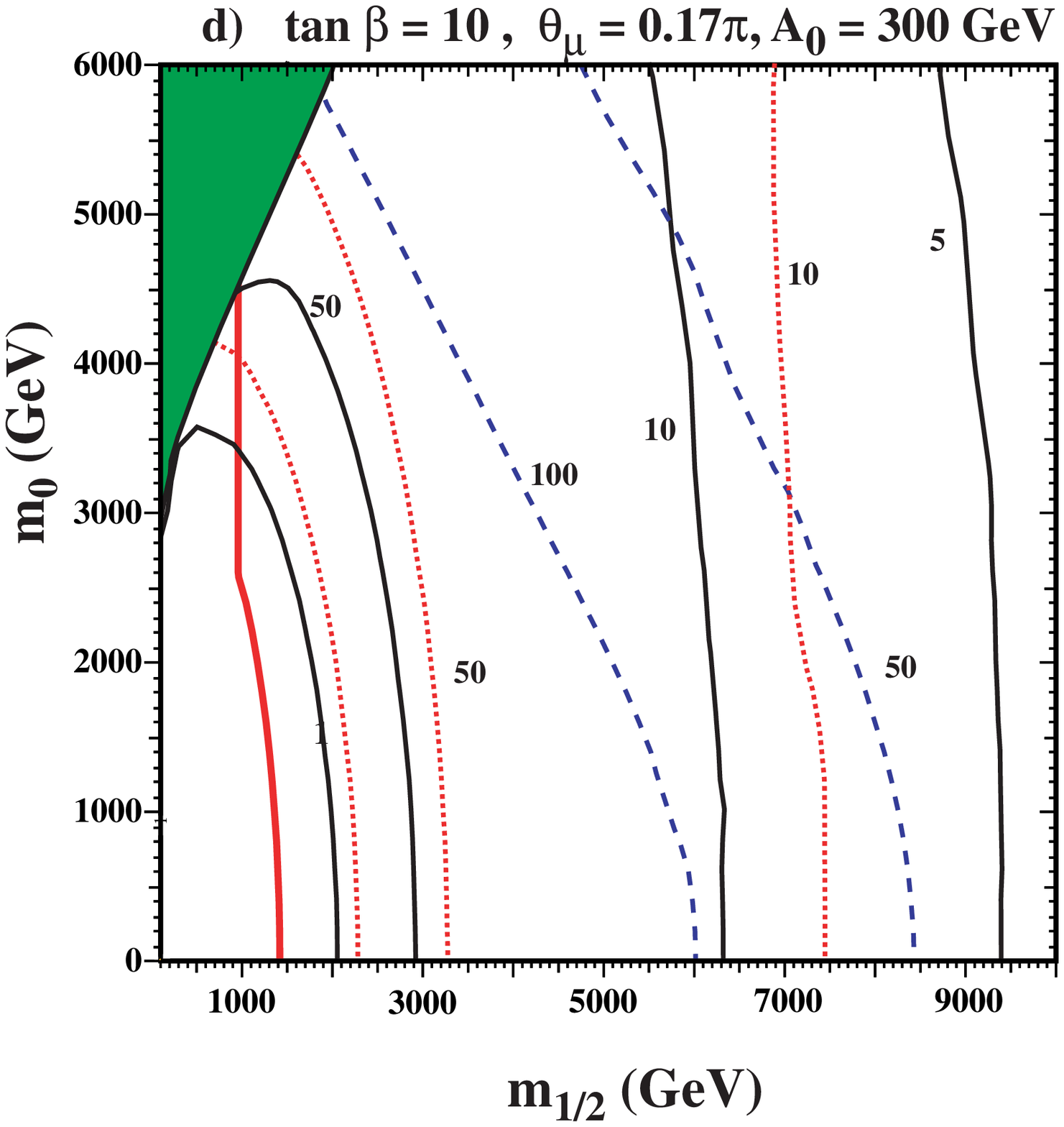}}
 \caption{\footnotesize As in Fig. \protect\ref{tb10}, the Tl (blue dashed), neutron (red dotted), and Hg
 (green solid) EDMs relative to their current experimental limits (a,b) and 
 the electron (blue dashed), the neutron (red dotted), and the deuteron (black solid)
 EDMs relative to possible future experimental limits (c,d) in the 
 $m_{1/2}, m_0$ plane of the CMSSM for $\tan \beta = 10$. In each panel, $|A_0| = 300$ GeV.
 In a and c), $\theta_A = 0.17\pi$,
 and in b) and d), $\theta_\mu = 0.17\pi$.}
\label{ext} 
\end{figure}

Our results are shown in  Fig.~\ref{ext}, where the current (a and b) 
and future (c and d) contours are shown for both $\theta_A$ (a and c) and $\theta_\mu$ (b and d).
Here we show only the value of the EDM relative to its experimental limit and suppress the other 
cosmological and phenomenological constraints.
Where possible, contours of 0, 1, 5, 10, 50, and 100 are shown.  
Note that the shaded region in the upper left corner of each
panel above the solid black curve is not viable in the CMSSM as there are 
no solutions which allow for radiative electroweak symmetry breaking.
(The boundary zone near this line is often called the focus point region,
or the hyperbolic branch.)
When $\theta_A = 0.17 \pi$ and $\theta_\mu = 0$, we see from panel a), that
the reach is not terribly strong.  For comparison, we show the expected reach
of the LHC based on its ability to detect colored sparticles with masses up
to 2 TeV \cite{baer, eoss7}.  However, we observe that the future sensitivity of
the neutron and deuteron EDM experiments will extend beyond the reach of the LHC
for $m_0 \la 4.5$ TeV. While it is interesting that the cancellation line apparently weakens the
constraints from paramagnetic sources, we emphasize that corrections from four-fermion sources
may change this significantly in the case of PbO and YbF, and that this cancellation may 
not survive in more general variants of the MSSM.

In contrast, when $\theta_\mu = 0.17 \pi$, the current reach of all three
EDM experiments already exceeds the projected reach of the LHC as seen
in panel b) of Fig. \ref{ext}.   For completeness, we show in panel d) the expected reach
from future EDM experiments.  This figure shows that for large $\th_\mu$ 
the EDMs can be sensitive to superpartner mass scales as high as a 
few tens of TeV. While such scenarios are apparently highly tuned, one may try to rescue naturalness
via SUSY breaking mechanisms, outside the CMSSM framework, where only the first two generation 
scalars are pushed above 10 TeV in order to cure the SUSY flavor problem by `brute force', while 
keeping the 3rd generation scalars light in order not to exacerbate the fine-tuning problem 
in the Higgs sector \cite{Nath:dn}. It is clear that the next generation
EDM searches will directly probe such scenarios.

The shape of these contours is easily understood given the large RG-induced admixture of $m_{1/2}$
to the squark masses. We see that the contours for the hadronic EDMs are squashed towards the
$m_0$ axis as a consequence of the rapid suppression of the EDMs as $m_{1/2}$ becomes large. This
effect is absent for the paramagnetic EDM bounds, and we observe a more uniform shape for these 
contours.

\section{Concluding Remarks}

We have performed a general analysis of the EDM constraints on the CMSSM parameter space,
combining a number of weak-scale threshold contributions to the fermion EDMs and CEDMs, and
also four-fermion operators. The latter contributions are generally subleading, but become
important at large $\tan\beta$ and must be taken into account for a complete analysis of
this regime. We have emphasized two main points as part of this analysis. Firstly, as
the experimental situation is rapidly evolving, we have presented estimates for the sensitivity
of the next generation of EDM experiments in each of the three sectors which currently provide the
strongest constraints; namely paramagnetic sources sensitive to the electron EDM, and the 
neutron and deuteron which are sensitive to (different) combinations of quark/gluon operators.
The most striking impact of the expected levels of improvement in sensitivity is that
the reach of EDMs induced by $\th_A$ increases significantly, to the extent that it exceeds that
of the LHC for large phases. The reach in terms of $\th_\mu$ already probes superpartner mass scales
that cannot be directly accessed at the LHC.

The second point, of technical importance, that we have emphasized is that 2-loop RG-induced corrections
to Im$M_i$ can significantly enhance the dependence of the electron EDM to $\th_A$, particularly so 
at large $\tan\beta$. While on first inspection it would appear that 
these RG-induced contributions generically dominate the Tl EDM by a factor of a few, it turns out 
that the relative signs are such that significant cancellations occur, and the bounds are often 
weakened, at least for moderate $\tan\beta$. For large $\tan\beta$, a number of different contributions 
become competitive, but on closer inspection one may observe by comparing Fig.~\ref{hightb}c with f 
that the constraints from Tl do in fact become stronger with the inclusion of the RG-induced 
corrections, although overall they remain relatively weak in this regime. A summary of some
of the constraints on the phases in specific benchmark scenarios is shown in Table~1.

A natural question is whether these conclusions will hold in more general examples of the 
MSSM, where a number of the universality assumptions are lifted, and the number of independent phases 
is correspondingly increased. While the most recent explorations of this 
kind \cite{Barger:2001nu,Abel:2001mc,dlopr} suggest that the picture is not dramatically 
modified, this is clearly a question worthy of further study.

\subsection*{Acknowledgements}
We thank A. Pilaftsis for drawing our attention to the threshold corrections in \cite{Pil99b}, and for helpful
related discussions. We also thank S. Huber and O. Vives for helpful discussions. 
The work of K.A.O. was supported in part 
by DOE grant DE--FG02--94ER--40823. The work of M.P. and Y.S. was supported in part by the NSERC of 
Canada, and Y.S.thanks the Perimeter Institute for its hospitality.

\end{document}